\documentclass[11pt,a4paper,pdftex]{article}
\pdfoutput=1
\usepackage{jcappub}
\usepackage{color}

\newcommand{\be}{\begin{eqnarray}}
\newcommand{\ee}{\end{eqnarray}}
\newcommand{\rar}{\rightarrow}

\title{Using iron line reverberation and spectroscopy to distinguish Kerr and non-Kerr black holes}

\author[a]{Jiachen Jiang,}
\author[a,1]{Cosimo Bambi,%
\note{Corresponding author}}
\author[b,2]{and James F. Steiner
\note{Hubble Fellow}}

\affiliation[a]{Center for Field Theory and Particle Physics and Department of Physics,\\
Fudan University, 220 Handan Road, 200433 Shanghai, China}
\affiliation[b]{Harvard-Smithsonian Center for Astrophysics,\\
60 Garden Street, Cambridge, MA 02138, United States}

\emailAdd{jcjiang12@fudan.edu.cn}
\emailAdd{bambi@fudan.edu.cn}
\emailAdd{jsteiner@cfa.harvard.edu}

\abstract{The iron K$\alpha$ line commonly observed in the X-ray spectrum of 
both stellar-mass and supermassive black hole candidates is produced by the 
illumination of a cold accretion disk by a hot corona. In this framework, the activation 
of a new flaring region in the hot corona imprints a time variation on the iron line 
spectrum. Future X-ray facilities with high time resolution and large effective 
areas may be able to measure the so-called 2-dimensional transfer function; 
that is, the iron line profile detected by a distant observer as a function of time in 
response to an instantaneous flare from the X-ray primary source. 
This work is a preliminary study to determine if and how such a
technique can provide more information about the spacetime geometry around
the compact object than the already possible measurements of the time-integrated
iron line profile. 
Within our simplified model, we find that a measurement of iron 
line reverberation can improve constraints appreciably given a sufficiently strong signal, though
that most of the information is present in the time-integrated spectrum.  Our aim is to 
test the Kerr metric. We find that current X-ray facilities and data are unable to provide
strong tests of the Kerr nature of supermassive black hole candidates.  
We consider an optimistic case of $10^5$ 
iron line photons from a next-generation data set.  
With such data, the reverberation model improves upon the spectral constraint by 
an order of magnitude.  }

\keywords{astrophysical black holes, modified gravity, X-rays}

\begin{document}

\maketitle


\section{Introduction}

In 4-dimensional general relativity, a black hole (BH) is a relatively simple object.
Neglecting a possible non-vanishing electric charge, the spacetime geometry is
described by the Kerr solution and the object is completely specified by the value
of its mass $M$ and of its spin parameter $a_* = a/M = J/M^2$, where $J$ is the BH spin
angular momentum~\cite{h1,h2,h3}. If we know $M$ and $a_*$, all the properties
of the spacetime can be calculated from these two quantities. Astrophysical BH
candidates are grouped into two classes:  (1) supermassive compact bodies in active
galactic nuclei (AGN), hosting masses $M \sim 10^5 - 10^{10}$~$M_\odot$; and (2) 
compact objects in X-ray binary systems with masses $M \approx 5 - 20$~$M_\odot$~\cite{nara}.
Both object classes are widely accepted to be BHs; on this basis, the spacetime 
geometry around them should therefore be well described by the Kerr solution. 
Indeed, initial deviations from the Kerr metric could be quickly radiated away through 
the emission of gravitational waves~\cite{p1,p2}. Any initial non-vanishing electric 
charge would be quickly neutralized because of the highly ionized host environment 
of these objects~\cite{bdp}. The deviation induced by the presence of an accretion 
disk is usually completely negligible, because the disk mass is many orders of 
magnitude smaller than the mass of the BH candidate~\cite{massivedisk}.

However, current observations cannot unambiguously confirm the Kerr BH paradigm.
For the time being, our principal insight is through robust measurements of the masses 
of these objects. Such measurements are obtained through dynamical studies
which employ the orbital motion of nearby or companion stars.  From these 
measurements, one can conclude that stellar-mass BH candidates in X-ray binary 
systems are too massive to be neutron or quark stars for any plausible matter 
equation of state~\cite{ozel1, ozel2, bh1,bh2}, while supermassive BH candidates at
the center of galaxies are surely too heavy, compact, and old to be clusters of neutron 
stars, since the expected cluster lifetime due to evaporation and physical collisions 
would be shorter than the age of these systems~\cite{bh3}. The non-observation of 
thermal radiation emitted by the possible surface of these objects is usually interpreted 
as evidence for the existence of an event (or at least of an apparent) horizon~\cite{bh4,bh5}. 
In the end, these considerations lead one to conclude that astrophysical BH candidates 
are most readily explained as Kerr BHs in the framework of conventional physics, and 
that in order for them to be something else would only be natural in presence of new 
physics.

Nevertheless, general relativity has been tested only for weak gravitational fields,
while the theory is almost unexplored in the limit of strong gravity~\cite{will}.
Theoretical arguments in support of the Kerr BH hypothesis are thus not completely
satisfactory, and one would like to find observational tests of whether the spacetime
geometry around these objects is really described by the Kerr metric, 
especially in light of recent novel considerations which suggest the 
possibility of macroscopic deviations from classical predictions~\cite{gia1,gia2,giddings}.
The original idea of testing the Kerr nature of astrophysical BH candidates was put 
forward about 20~years ago in Ref.~\cite{gw}, where a test was proposed using the 
gravitational waves emitted by a system of a stellar-mass compact object orbiting 
around a supermassive BH candidate. This testing ground was further explored by 
other authors~\cite{gw1,gw2,gw3}. More recently, there have been significant
efforts examining possible tests of the Kerr BH hypothesis from the properties of 
the electromagnetic radiation emitted by the accretion disks of these systems.  Such 
tests have been proposed using both present X-ray data as well as future observations 
in the X-ray, NIR, sub-mm, and radio bands~\cite{cfm1,cfm2,cfm3,cfm4, iron1,iron2,iron3, qpo1,qpo2,qpo3, cfm-iron, sh1,sh2,sh3,sh4,pulsar,hot2,2-aaa}. For a review on the 
subject, see e.g.~\cite{rev1,rev2} and references therein.

Today, there are two techniques commonly employed in probing the spacetime 
geometry around astrophysical BH candidates; that is, the continuum-fitting 
method~\cite{k1,k2,k3,k3b} and the analysis of the iron K$\alpha$ line~\cite{k4,k5,k6,k6b}. 
These two techniques have been developed to estimate the spin parameter of 
BH candidates under the assumption that the spacetime geometry around them 
is described by the Kerr metric. Both also rely on the assumption that the inner-edge 
of the accretion disk is located at the innermost stable circular orbit 
(ISCO)\footnote{Here it is clearly important to select the sources in which the inner edge of the disk is at the ISCO radius. This is likely the case when the accretion luminosity is between a few percent and about 30\% the Eddington luminosity, which is a selection criterion commonly employed in the continuum-fitting method~\cite{k3b,ste1,vnnn-1,vnnn-2}.  We note that when advection becomes important the inner edge of the disk deviates predictably from the ISCO.}, 
an assertion which is grounded in the observed constancy of accretion-disk inner 
radii~\cite{ste1}. 
The extension to non-Kerr backgrounds to test the Kerr nature of 
BH candidates is straightforward~\cite{cfm1,cfm2,cfm3,cfm4,iron1,iron2,iron3}. 
Recently, there has been an increasing interest in methods exploiting the time domain, in particular in the relativistic precession model~\cite{v4-m1,v4-m2} and eclipse mapping~\cite{v4-r,v4-matt}. Even these techniques can be potentially used to test the Kerr metric~\cite{v4-cb}.

To test the nature of BH candidates we can use an approach similar to the 
Parametrized Post-Newtonian (PPN) formalism of Solar System 
experiments~\cite{will}. One assumes a general background metric with some 
free parameters to be determined by observations. In the case of general relativity, 
the exact value of these parameters is known and one wants to check if the 
measurements of these parameters is consistent with the values required by 
general relativity. Unfortunately, at present there is not a framework as general
as the PPN formalism to test the Kerr BH hypothesis. One can anyway consider a 
BH spacetime in which the compact object is specified by a mass $M$, a spin 
parameter $a_*$, and at least one deformation parameter. The latter measures 
possible deviations from the Kerr solution, which is recovered when the deformation 
parameter vanishes. One can then try to constrain the deformation parameter and 
see if observations require a non-Kerr element or not.

In general, however, it is not easy to constrain the deformation parameter. Critically, 
the continuum-fitting and iron line measurements seem able to measure only one 
parameter of the spacetime geometry close to BH candidates.  In other words, there 
is a strong degeneracy between any constraint on spin and deformation parameters,
and therefore a single measurement is only able to constrain a combination of these 
two quantities. By measuring different relativistic effects using a variety of techniques 
for the same source, one can potentially break such degeneracy~\cite{j1,j2,hot1,naoki}.
The current data fall short of providing strong independent constraints, but it will be 
hopefully achievable in the future. With current observations, we can at most exclude 
some very exotic BH alternatives, like compact objects without event horizon~\cite{cc1} 
and some classes of wormholes~\cite{cc2}. In the case of objects that look like
very-fast rotating Kerr BHs, it is generally possible to obtain a bounded constraint 
ellipse along the spin--deformation parameter plane~\cite{cc3}; usually it is difficult
for deformations and low spin to mimic a Kerr BH with spin parameter close to 1 
(but see Ref.~\cite{cpr-metric}).

In this Paper, we aim to study the information provided by iron line reverberation to 
test the Kerr nature of BH candidates. The exact origin and geometry of the corona 
that produces the iron line is not known, 
although some progresses have been achieved~\cite{v4-wg}, 
but the line exhibits a high degree of variability 
on short timescales which seems to be associated with the activation of new flaring 
regions.  This, in turn, spawns temporal variations in the iron line, owing to the different
propagation time for different photon paths. Because of the limited count rates in the 
iron line with current X-ray facilities, present observations are integrated over many 
thousands of seconds.  This causes a loss of information. Future X-ray facilities with 
large effective areas may be able to study the temporal change in response to the 
activation of new 
flares\footnote{We note that the corona variability is unlikely described by actual ``flares''. There is indeed evidence that the emission of different regions is correlated on all time scales~\cite{v4-uttley}. Nevertheless, such terminology is commonplace, and we employ the flare and response picture for its simplicity.}. 
Reverberation of the line can be exploited to gain insight into 
the system~\cite{r1,r2,r3}. Here, we aim to figure out the advantage of the study of line
reverberation to constrain possible deviations from the Kerr metric around BH candidates. 
The key-quantity to study is the 2-dimensional (2D) transfer function, which essentially
corresponds to the time dependent iron line profile produced by an X-ray source emitting 
an instantaneous flare. While the advantage of reverberation observations over pure 
iron line spectroscopy depends on the quality of the available data (i.e., on the specific 
features of the detector, the source, and the observation), we find that high-quality 
reverberation measurements 
can improve upon constraints from 
the spectrum alone, even though most information is in the time-integrated 
profile. The key to using reverberation for testing the Kerr paradigm is the accumulation  of 
sufficient line photons. Present setups surely oversimplify the picture. 
With current X-ray facilities, a good observation of a supermassive black hole 
can have  $\sim 10^3$ photons 
in the iron line.  With such a signal, even very deformed 
objects may be interpreted as Kerr BHs with a different spin. We instead consider an
optimistic $10^5$ photon data set, with a next-generation X-ray satellite in mind. With such
data, we can perform a discerning test to check whether BH candidates are in fact the Kerr BHs 
of general relativity. Reverberation data can sharpen this test substantially.

The content of the Paper is as follows. In Section~\ref{s-rev}, we review the approach 
by which reverberation of the iron line is used to infer the geometry of the X-ray source 
and the BH spin in a Kerr background. In 
Section~\ref{s-bh}, we extend the formalism to non-Kerr backgrounds and show 
how reverberation can be used to test the actual nature of an astrophysical BH 
candidate. Section~\ref{s-sim} demonstrates a preliminary quantitative analysis in 
which we compare iron line profiles and 2D transfer functions in Kerr and non-Kerr 
spacetimes to examine the possibility of distinguishing Kerr and non-Kerr BHs. 
Section~\ref{s-d} is devoted to a discussion of our results, with
some attention to using iron line reverberation to test the Kerr metric 
and also the simplifications of our model that should be improved upon
in future studies.
Our summary and conclusions are reported in Section~\ref{s-c}. Throughout the Paper, 
we use units in which $G_{\rm N} = c = 1$, so lengths and times are measured in units 
of $M$, the mass of the BH.

\section{Iron line reverberation \label{s-rev}}

\subsection{The corona-disk model in lamppost geometry}

Optically-bright AGNs are thought to host a central
supermassive BH surrounded by an optically thick and geometrically
thin accretion disk. In the disk, the mass flow is directed inward and
the angular momentum flow is directed outward; the accretion disk
radiates like a blackbody locally, or as a multi-color blackbody when
integrated radially. In addition to this blackbody-like ``thermal''
component, the electromagnetic spectrum has other features which
result from a hotter, usually optically-thin, electron cloud termed
the ``corona'' which enshrouds the central disk and acts as a source
of X-rays. This X-ray emitting corona is often approximated as a point
source located on the axis of the accretion disk and just above the
BH\footnote{Although any true system must be physically extended and
accordingly more complex, here we adopt the usual on-axis isotropic point
source paradigm, which is theoretically clear and simple.}. This
arrangement is often referred to as a ``lamppost geometry'' and the
full structure is termed the ``corona-disk
model''~\cite{l1,l2}. 
The lamppost geometry requires a plasma of electrons very close to the BH and such a set-up may be realized in the case of the base of a jet. However, other geometries are possible, and an example is the family of ``sandwich models''~\cite{v4-sw1,v4-sw2}. Geometries different from the lamppost one are difficult to model using a fully general relativistic approach, because the number of free parameters is higher and the structure still ambiguous. However, some work has been done, see e.g.~\cite{v4-sw3}. Here to illustrate these effects we consider the simplest case of lamppost geometry.

Different corona-disk models, consisting of a
Kerr BH, an accretion disk, and an emitting source, have 2
characteristic parameters to be described. One is the spin parameter
$a_*$, with $|a_*| \le 1$ in order to describe a BH and not a naked
singularity. The second one is the height of the source above the
accretion disk, $h$, which is important for measuring time-dependent
iron line signals. The standard framework to describe geometrically thin and optically 
thick accretion disks is the Novikov-Thorne model~\cite{nt-model, nt-model2}, in which
the disk is on the equatorial plane and the particles of the gas move
on nearly geodesic circular orbits (i.e., Keplerian motion).  
The accretion disk is expected to have its inner edge at the radius of the ISCO,
which in the Kerr background only depends on $a_*$ and ranges from
$r_{\rm ISCO} = 6\,M$ for a Schwarzschild BH to $r_{\rm ISCO} = M$ for
an extremal Kerr BH with $a_* = 1$ and a corotating disk.  In response
of the illumination of the disk by the hot corona, fluorescent
emission is produced at the disk's surface, termed a ``reflection
component'' which is most distinguished by its prominent emission
lines \cite{gfab, rofab, gar1}. Here we will focus our attention only on the iron K$\alpha$
line at $\sim6.4$~keV, which is the strongest reflection line from 
BH candidates. In what follows, we set the emissivity index
of the corona's intensity across the disk at $q = 3$ (unless stated
otherwise), i.e., the coronal flux received by the disk is directly
proportional to $r^{-q}$, where $r$ is the disk's radius.  We note
that this approach oversimplifies the problem, because $q
= 3$ is the asymptotic limit expected at large radii $r \gg M$, while
at small radii near $r \approx h$, a proper reflection, lamppost model produces
significant deviations~\cite{gar1, dauser}.  However, because our goal is to
present a qualitative exploration of Kerr vs. non-Kerr geometries and
given that the precise coronal geometry is uncertain,
we leave a more detailed treatment to future followup studies.

\begin{figure}
\begin{center}
\includegraphics[type=pdf,ext=.pdf,read=.pdf,width=7.0cm]{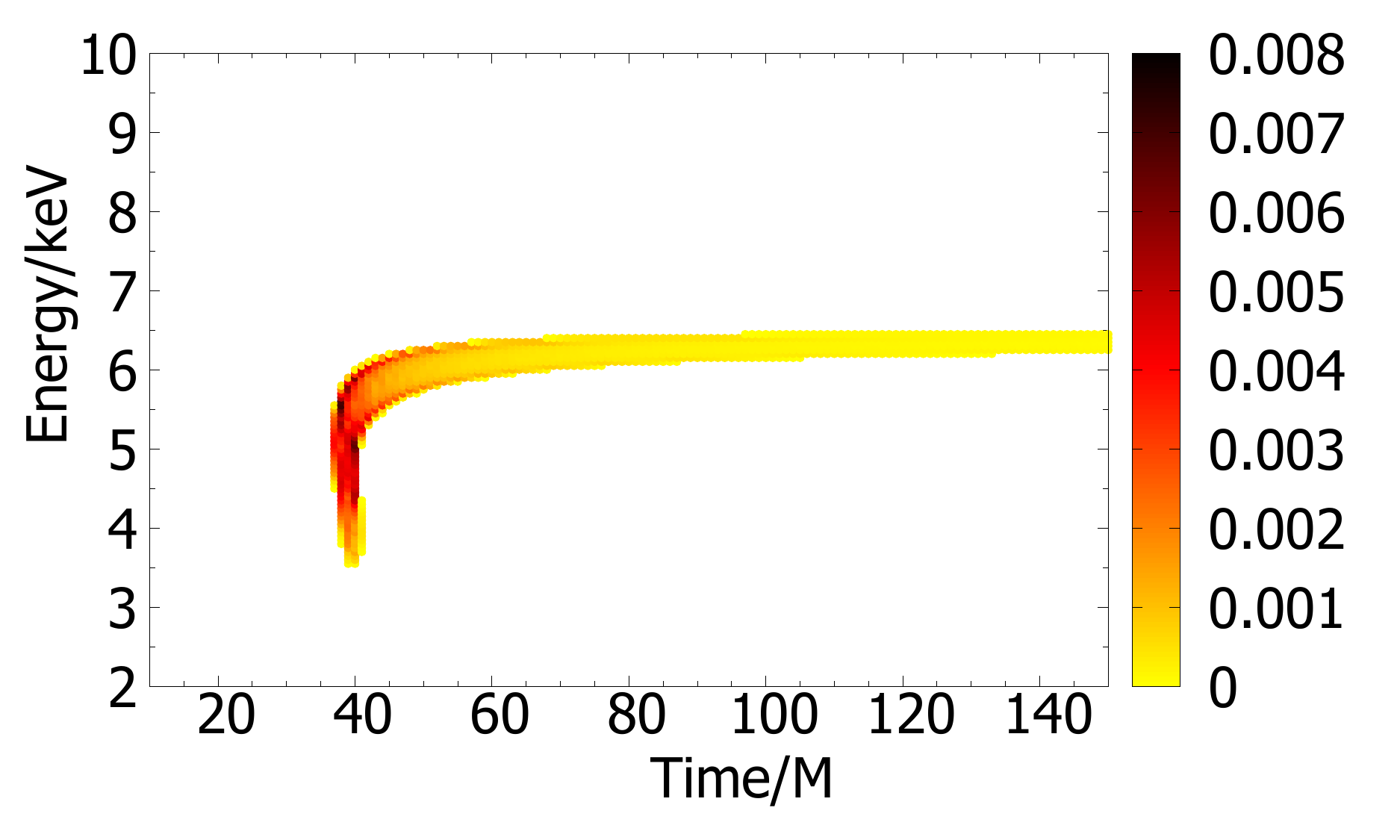}
\hspace{0.5cm}
\includegraphics[type=pdf,ext=.pdf,read=.pdf,width=7.0cm]{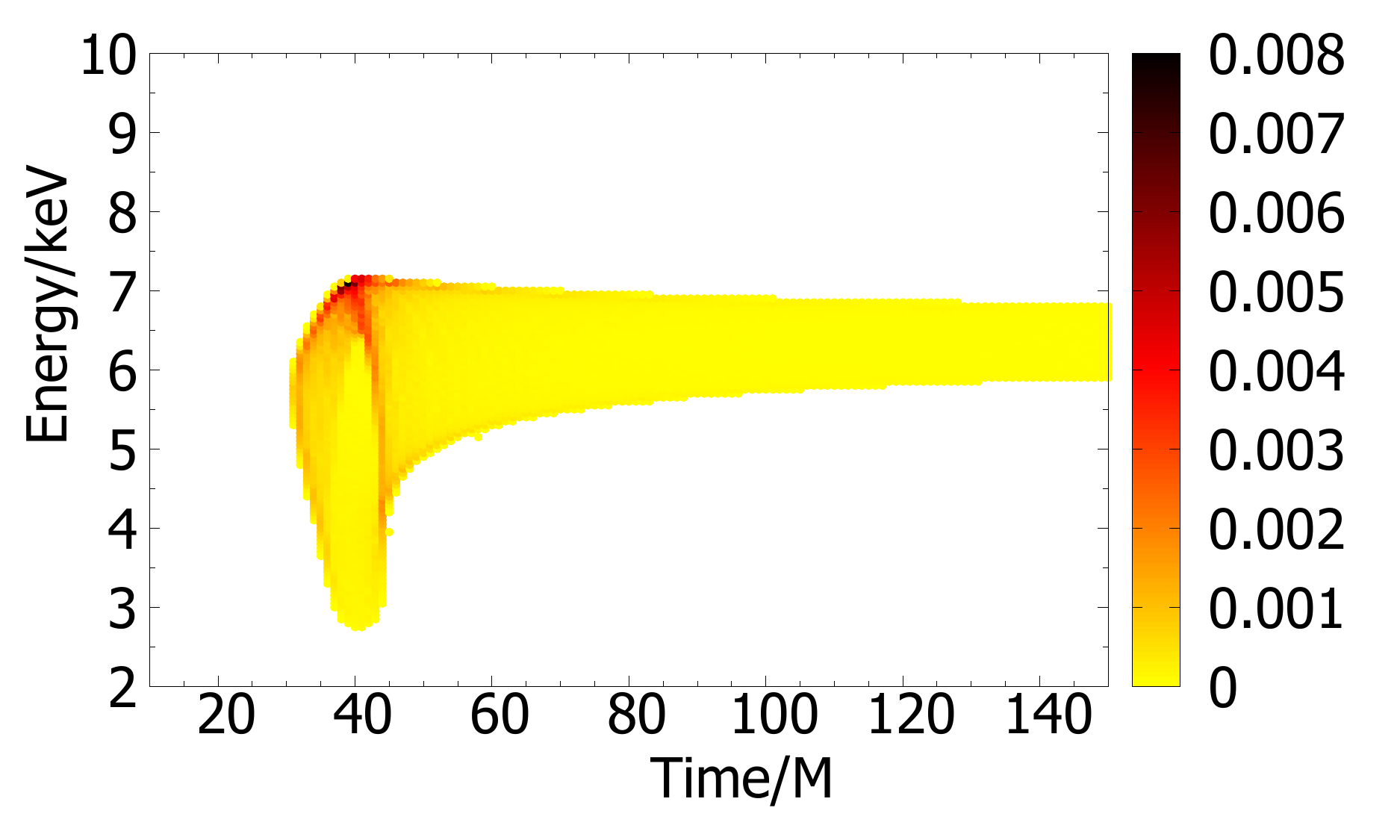} \\
\vspace{0.6cm}
\includegraphics[type=pdf,ext=.pdf,read=.pdf,width=7.0cm]{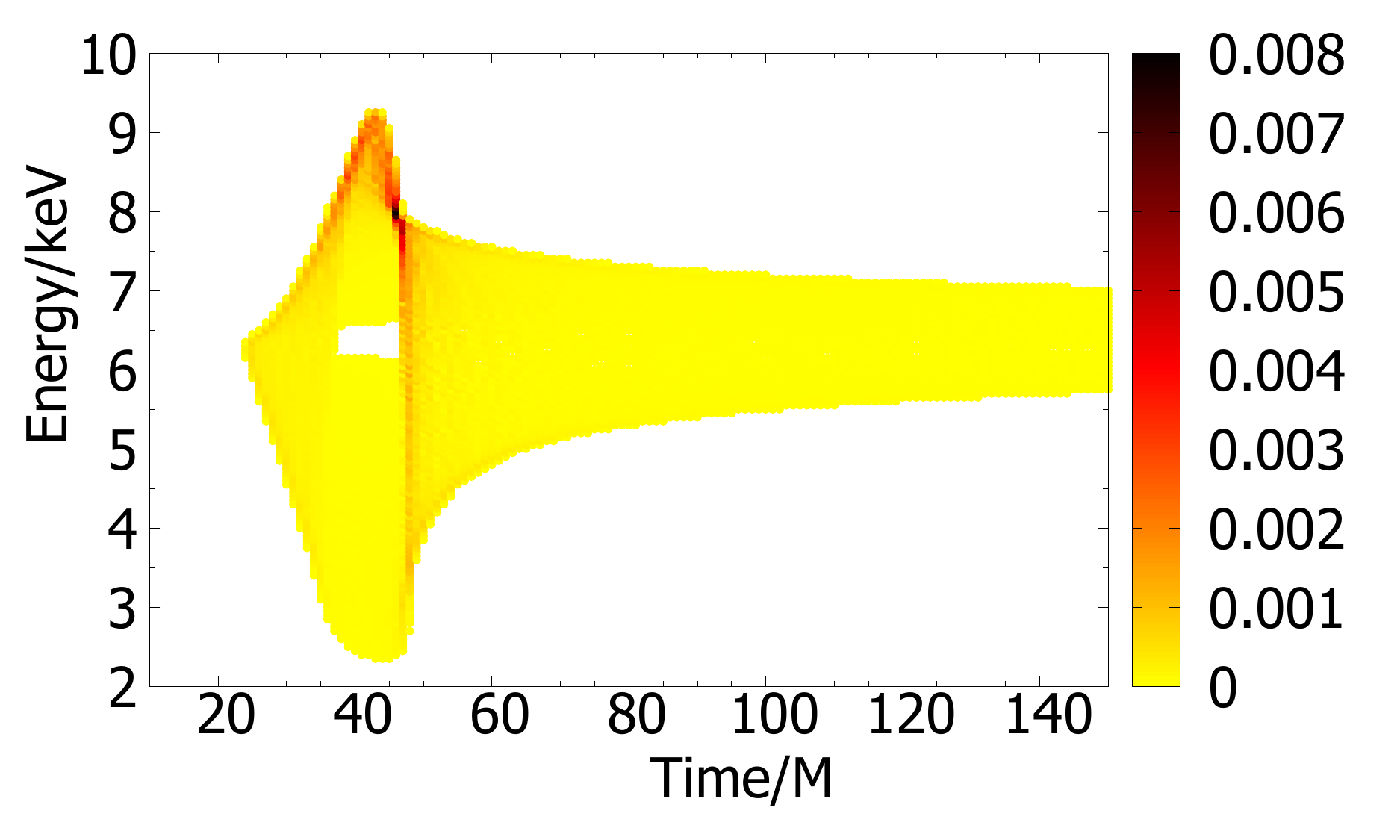}
\end{center}
\caption{Transfer functions for a Kerr BH with spin parameter $a_* = 0.5$.
The inclination angle is $i=10^\circ$ (top left panel), $i=45^\circ$ (top right panel),
and $i=80^\circ$ (bottom panel). The height of the source is $h = 10 \, M$ and
the index of the intensity profile is $q = 3$. 
Flux in arbitrary units. See the text for more details.}
\label{fig1}
\end{figure}

\subsection{Line reverberation from an on-axis coronal flare}

In the framework of the corona-disk model, ``reverberation'' refers to
the iron line signal as a function of time in response to a
$\delta$-function like pulse of radiation from the X-ray primary
source (i.e., the corona)\footnote{We neglect the timescale over which
  X-rays are reprocessed in the disk atmosphere.}. The resulting line
spectrum as a function of both time and across photon energy is called
the 2D transfer function.  As we show, the shape of this 2D transfer
function is related to fundamental properties of the BH and also the
system geometry.  The transfer function becomes nonzero as the first
 photons reach the observer.  Such photons are by construct 
those with the shortest path between the primary X-ray source, the
disk, and the distant observer. The radial coordinate at
which the shortest-path X-rays intercept the disk depends on the height of the
source $h$.   Likewise, the shortest path is dependent on the
inclination of the disk with respect to the observer's line of sight,
$i$.   All other photons reaching the detector hit the disk at smaller and larger
radii than that critical initial path, so they may have lower/higher energies as a result of a
stronger/weaker gravitational redshift and different Doppler
boosting. In the case of an almost face-on disk (small $i$), the
situation is less complicated because the Doppler boosting is small
and the gravitational redshift dominates. 
In this case, the transfer
function exhibits two distinct branches (see the top left panel
in Fig.~\ref{fig1}): the high energy branch represents photons coming
from further out in the disk, while the low energy
branch is produced
by strongly redshifted photons from the innermost disk.  The latter
track terminates when photons
from the ISCO reach the detector. At later times, the photon energy
tends toward the value 6.4~keV, the rest-frame energy of iron K$\alpha$ 
line, because at the larger radii at which these photons originate, the
gravitational redshift is of diminishing importance.

\begin{figure}
\begin{center}
\includegraphics[type=pdf,ext=.pdf,read=.pdf,width=7.0cm]{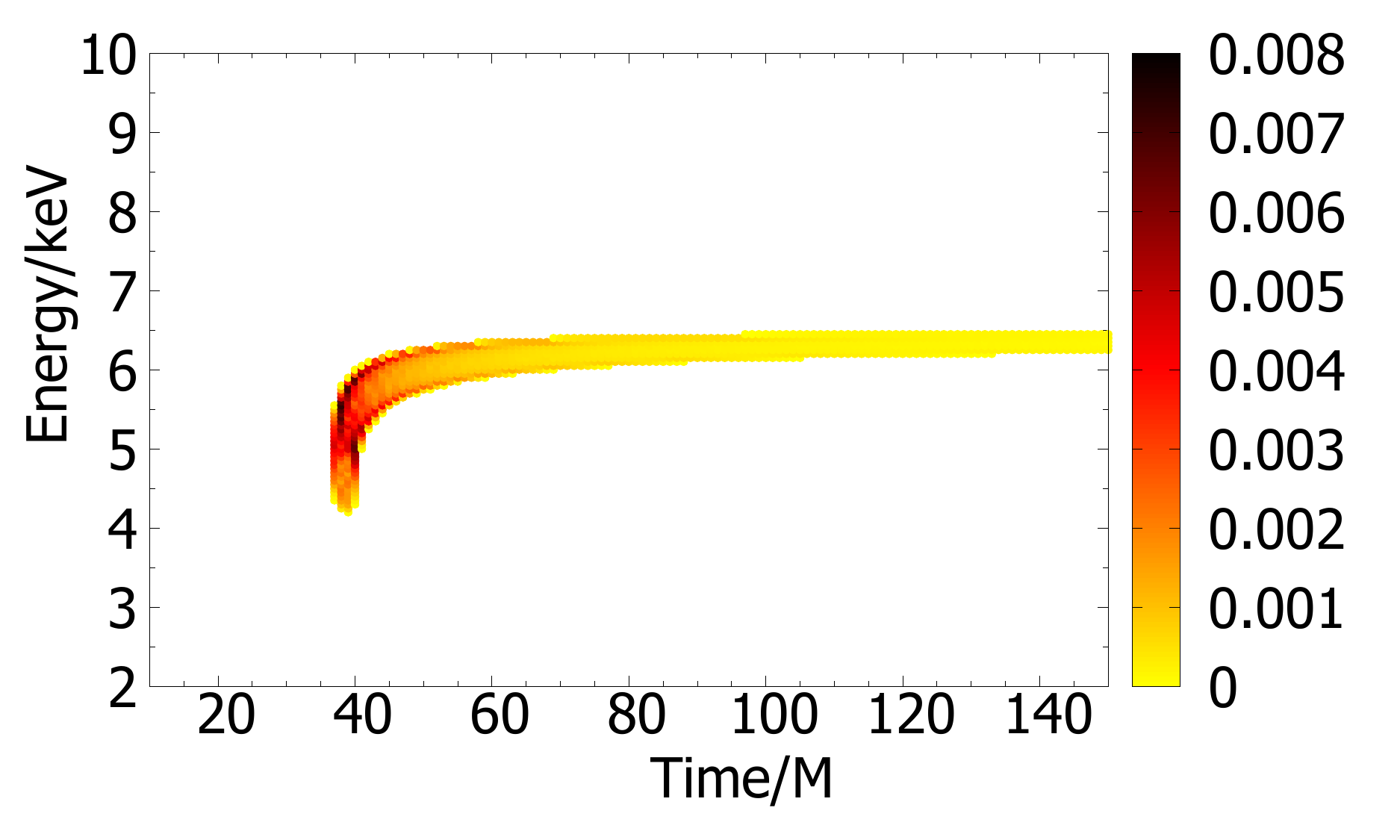}
\hspace{0.5cm}
\includegraphics[type=pdf,ext=.pdf,read=.pdf,width=7.0cm]{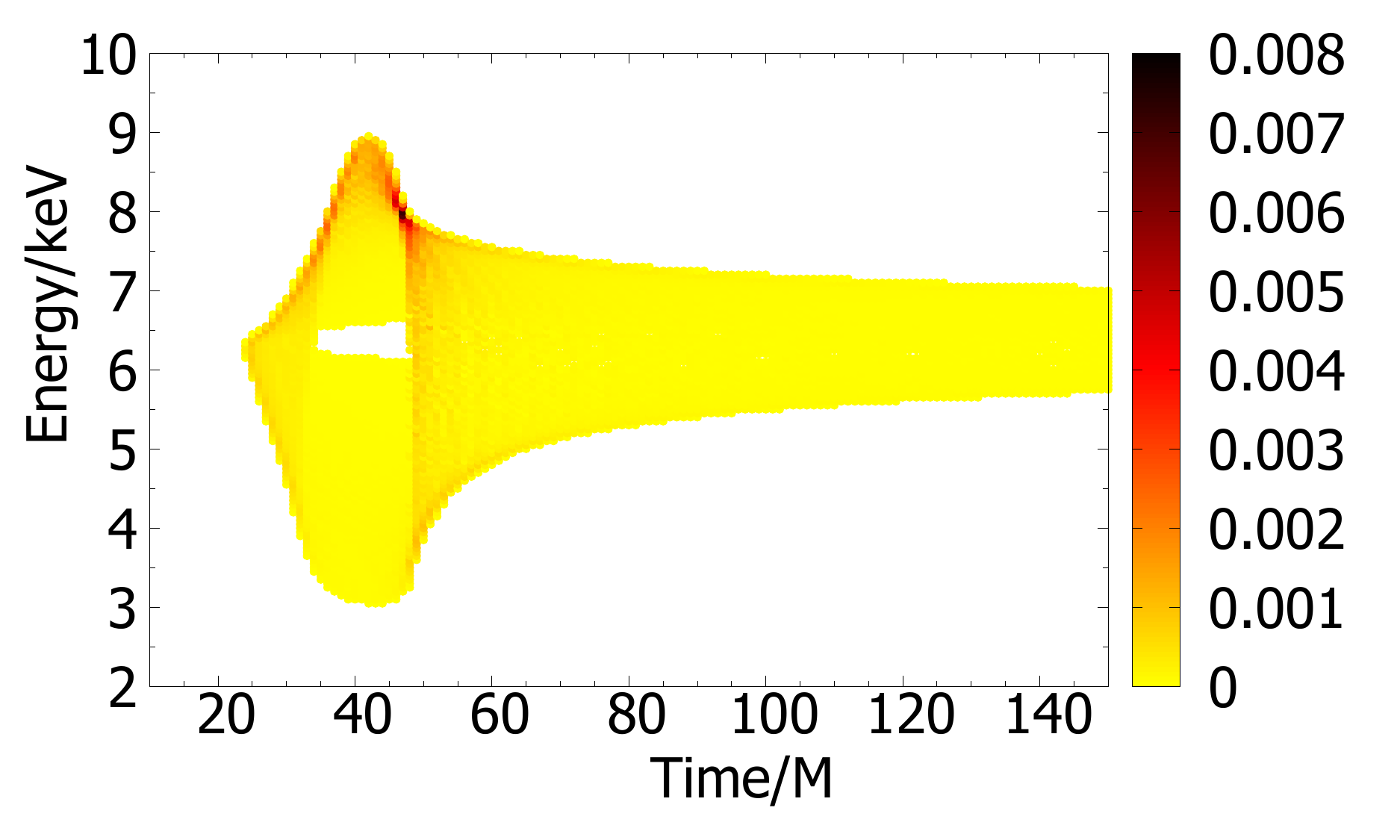}
\end{center}
\caption{Transfer functions for a Schwarzschild BH ($a_*=0$). The inclination
angle is $i=10^\circ$ (left panel) and $i=80^\circ$ (right panel). The height of the
source is $h = 10 \, M$ and the emissivity index of the intensity profile is $q = 3$. 
Flux in arbitrary units. See the text for more details.}
\label{fig2}
\vspace{0.6cm}
\begin{center}
\includegraphics[type=pdf,ext=.pdf,read=.pdf,width=7.0cm]{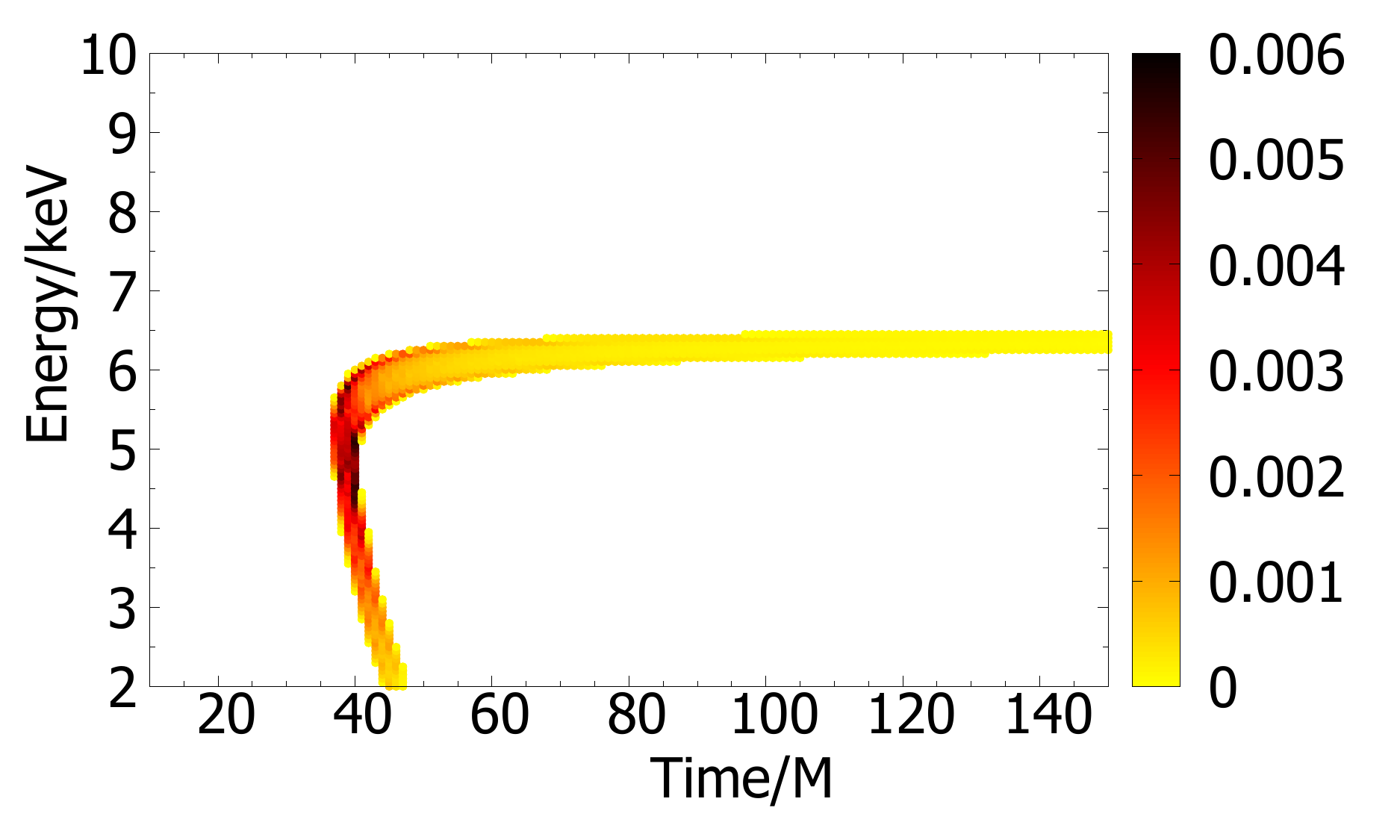}
\hspace{0.5cm}
\includegraphics[type=pdf,ext=.pdf,read=.pdf,width=7.0cm]{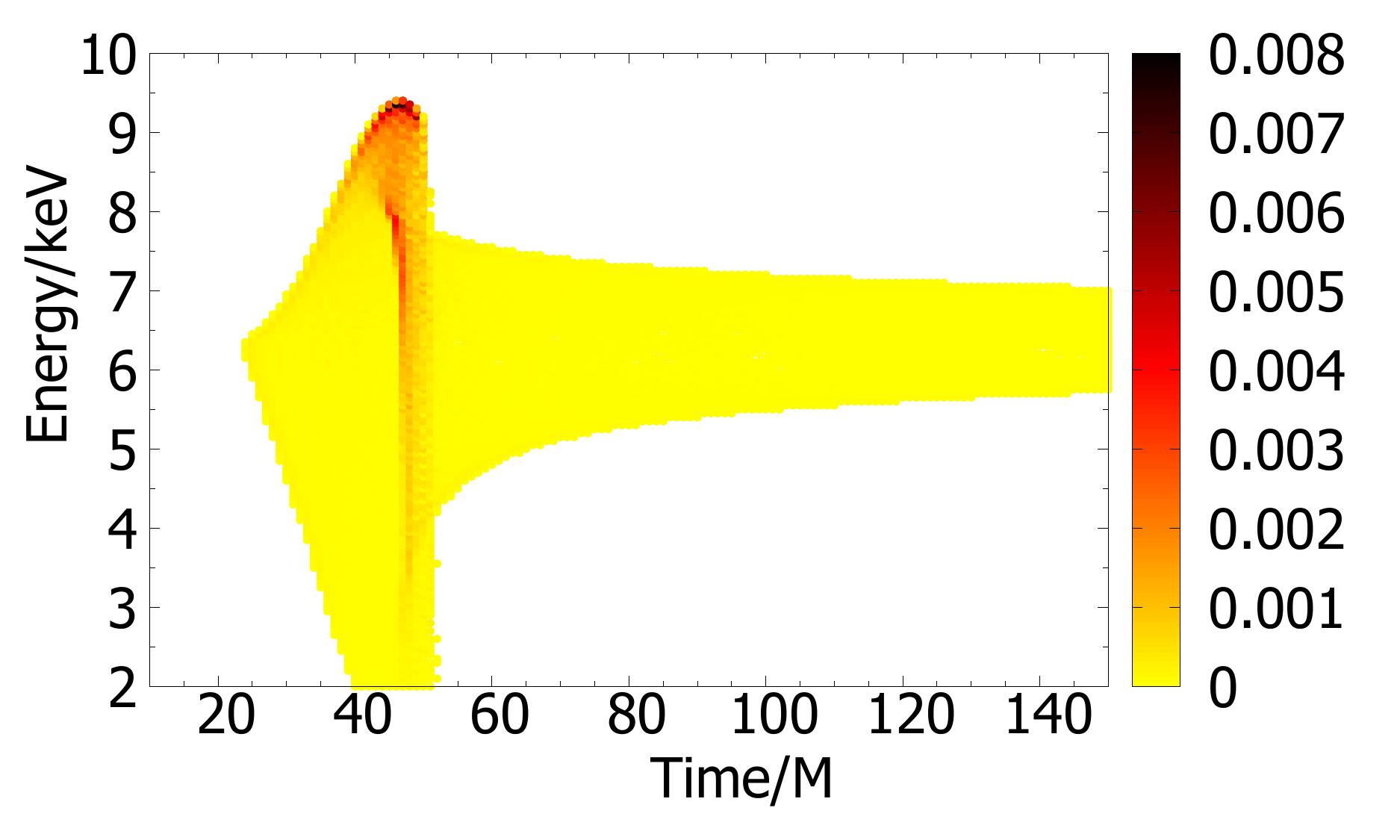}
\end{center}
\caption{Transfer functions for a Kerr BH with spin parameter $a_* = 0.95$.
The inclination angle is $i=10^\circ$ (left panel) and $i=80^\circ$ (right panel).
The height of the source is $h = 10 \, M$ and the index of the intensity profile
is $q = 3$. Flux in arbitrary units.
See the text for more details.}
\label{fig3}
\vspace{0.6cm}
\begin{center}
\includegraphics[type=pdf,ext=.pdf,read=.pdf,width=7.0cm]{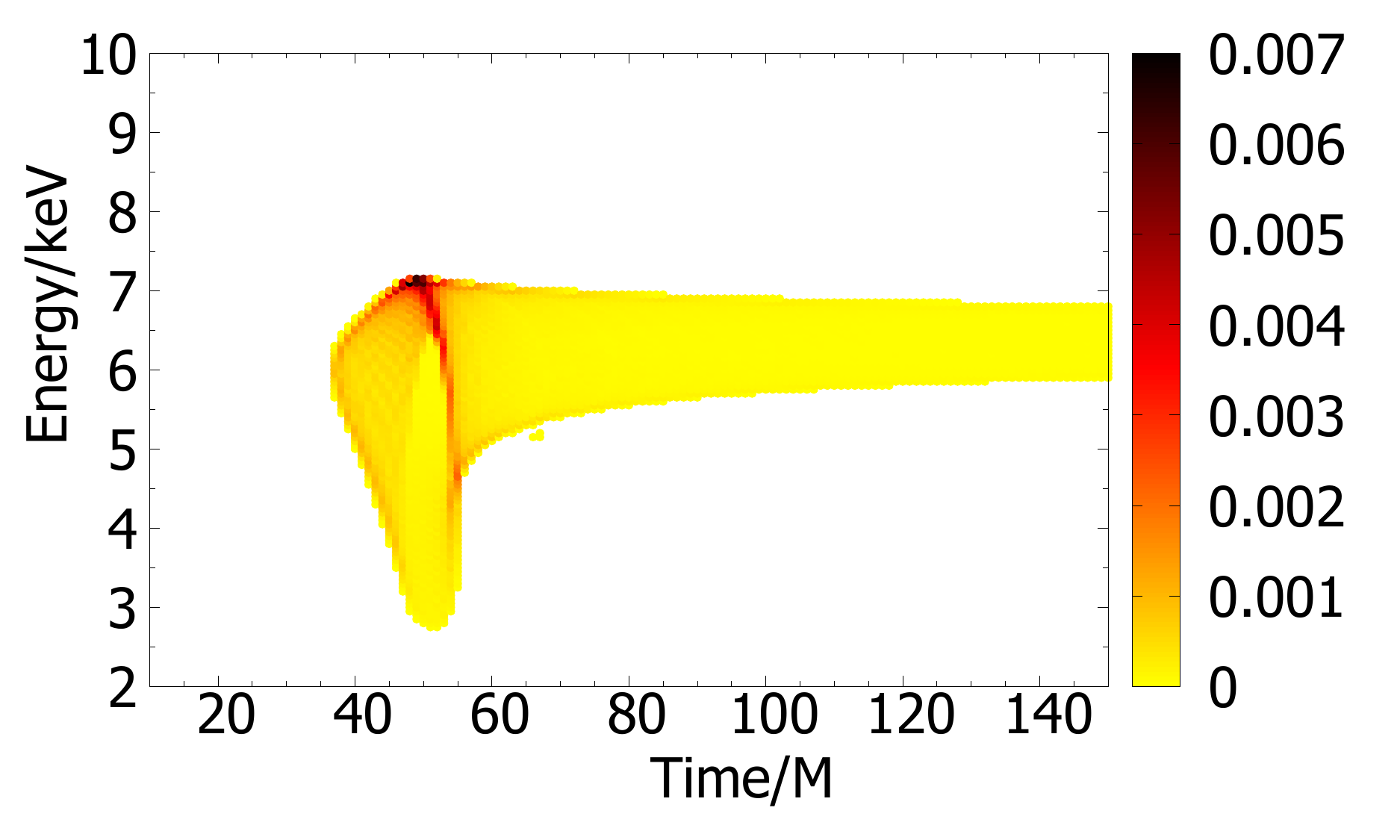}
\hspace{0.5cm}
\includegraphics[type=pdf,ext=.pdf,read=.pdf,width=7.0cm]{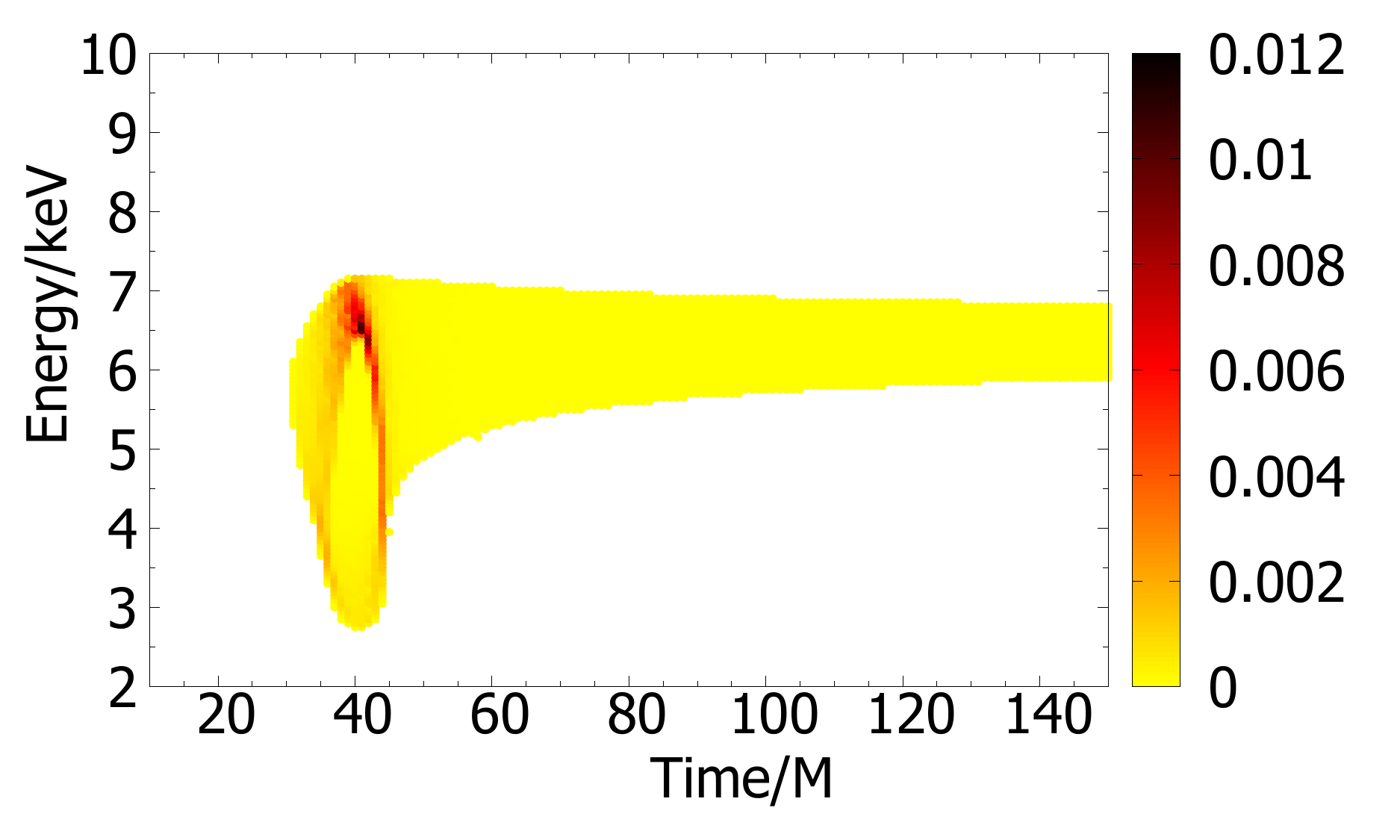}
\end{center}
\caption{Transfer functions for a Kerr BH with spin parameter $a_* = 0.5$
and an inclination angle $i=45^\circ$. In the left panel, the height of the source
is $h = 20 \, M$ and the index of the intensity profile is $q = 3$. In the
right panel, the height of the source is $h = 10 \, M$ and the index of the intensity
profile is $q = 6$. Flux in arbitrary units. See the text for more details.}
\label{fig4}
\end{figure}

\subsection{The 2D transfer function for a Kerr background}

In this paper, the calculations are done with an extension of the code described in Refs.~\cite{cfm3,iron3}, which is based on a ray-tracing approach and photon trajectories are integrated backward in time from the plane of the distant observer to the emission point on the disk. The resulting 2D transfer functions for different viewing inclination angles are shown in
Fig.~\ref{fig1}. When the inclination angle of the observer is small, the Doppler
boosting is subdominant and the gravitational redshift determines the shape of the
transfer function. As the inclination angle increases, Doppler
boosting becomes increasingly important. For photons hitting the disk at radii 
$r \approx 10-15\,M$,
the Doppler blueshift can significantly exceed the gravitational redshift and
produces a characteristic high-energy peak in the transfer function.
Meanwhile, for photons hitting the disk
at even larger radii, both gravitational redshift and Doppler boosting
decrease in strength, with the net result that the photon energy is
lower.  The top left panel in Fig.~\ref{fig1}
shows the transfer function for a low viewing angle, namely $i = 10^\circ$, while
the other two panels show broadened transfer functions corresponding
to mid-range and nearly edge-on viewing
angles, respectively $i = 45^\circ$ and $i = 80^\circ$. The higher the
inclination angle, the more
complicated the transfer function becomes. However, at late times the
problem again simplifies, because only line emission from further out
persists, and there the gravitational redshift becomes negligible
and the width of the band around 6.4~keV is due to the Doppler redshift and
blueshift caused by the orbital motion of the gas comprising the
accretion disk.  In fact, 
the width of the transfer
function at late times has constraining power on the inclination angle
of the disk.  
We note that the disk in many AGN systems may be warped, and in such a case the inclination angle of the inner part of the disk would not be the same as the inclination angle of the disk at its outer edge. The scales at which a warp from inner domain to outer are expected to be an order of magnitude or more larger than those relevant to the reverberation maps in this work (e.g., \cite{king05}), which probe scales $\lesssim 100 M$, at which the Bardeen-Peterson effect is expected to keep the disk in strict alignment with the spin axis (but we also note there is still uncertainty in the true nature of AGN disks). At larger radii ($>>10M$), the background metric has negligible effect and indeed this
regime of the transfer function is essentially the same for Schwarzschild BHs and fast-rotating
Kerr BHs (Figs.~\ref{fig2} and \ref{fig3}). However, the signal becomes
fainter and fainter, so any measurement would accordingly become tenuous
in strength.

The transfer functions for BHs with different spin parameters in the
Kerr background differ at early
times, since those photons originate in the strong-gravity region closest to the BH. The effect of 
spin at low inclination angles can be seen in the top left panel of Fig.~\ref{fig1}
and in the left panels in Figs.~\ref{fig2} and \ref{fig3}. The
low-energy, ``red wing'' of the
transfer function depends significantly on the radius of the inner edge of the disk, which
in the case of a geometrically thin and optically thick disk is
thought to correspond with the ISCO
radius (e.g., \cite{zhu1, schnit1}). For a Schwarzschild BH, the ISCO radius 
is at $r_{\rm ISCO} = 6\, M$
and the minimum photon energy is about 4~keV for $i=10^\circ$.
As the spin parameter increases, the ISCO radius decreases and the minimum
photon energy decreases. For $a_* = 0.95$ and $i=10^\circ$, the minimum photon
energy of the emission line
is lower than 2~keV and is off-scale for the left panel in Fig.~\ref{fig3}.

The role of the source height is evident from the shift in the time of
peak response when comparing the top right panel in 
Fig.~\ref{fig1} and the left panel in Fig.~\ref{fig4}. The two transfer functions differ 
only in the corona's height; $h=10\,M$ for the former, and $h=20\,M$ for the latter.  
The source height affects the travel time of the photons from the
X-ray source to the disk. If $h$ increases, such a travel time increases as well,
and therefore the distant observer detects a longer delay between the primary 
X-ray component that comes directly from the X-ray source and the reflection
component from the disk. In the example shown in the left panel in Fig.~\ref{fig4} 
for a source with height $h=20\,M$ and a viewing angle $i = 45^\circ$, the transfer 
function begins at about $35\,M$, while the one in the right panel in Fig.~\ref{fig1} 
for $h=10\,M$ begins at about $30\,M$. The height of the source affects also the 
time difference of the illumination of gas in the accretion disk at different radii. 
In the case of an on-axis point source in flat spacetime, the travel time from the 
source to the disk monotonically increases with the disk radius and the time difference 
between two different radii decreases as the height of the source
increases,  tending toward zero as $h \rar \infty$. Because the
observer's viewing angle affects the photon travel 
time from disk to observer, both the time difference between primary 
and reflection components, and also the shape of the transfer function
depend non-trivially on the source height and viewing angle. However, the 
height of the source does not affect the photon energy, while the viewing angle 
does via the magnitude of Doppler boosting.

In flat spacetime, the reflected radiation from the accretion disk is given by
\be
I\propto \frac{h}{(r^2+h^2)^\frac{3}{2}} \, ,
\ee
and without any relativistic effects, the intensity is related to the radius. 
A constant emissivity index is thus a simplification. In actuality, different radii on 
the accretion disk will have different emissivity indices. For larger radii, the emissivity index 
tends toward $q = 3$. 
We explore the effect of changing the emissivity index from $q = 3$ to $q = 6$, and
note that the shape of the transfer function does not change at all, 
although the value of flux does, see the top right
panel in Fig.~\ref{fig1} and the right panel in Fig.~\ref{fig4}.
Darker colors represent higher flux values.  A steeper index causes
the flux to fall off more rapidly with radius, and thus the later
times at which those larger radii dominate the transfer function are
correspondingly fainter for $q=6$ than $q=3$.

\begin{figure}
\begin{center}
\includegraphics[type=pdf,ext=.pdf,read=.pdf,width=7.0cm]{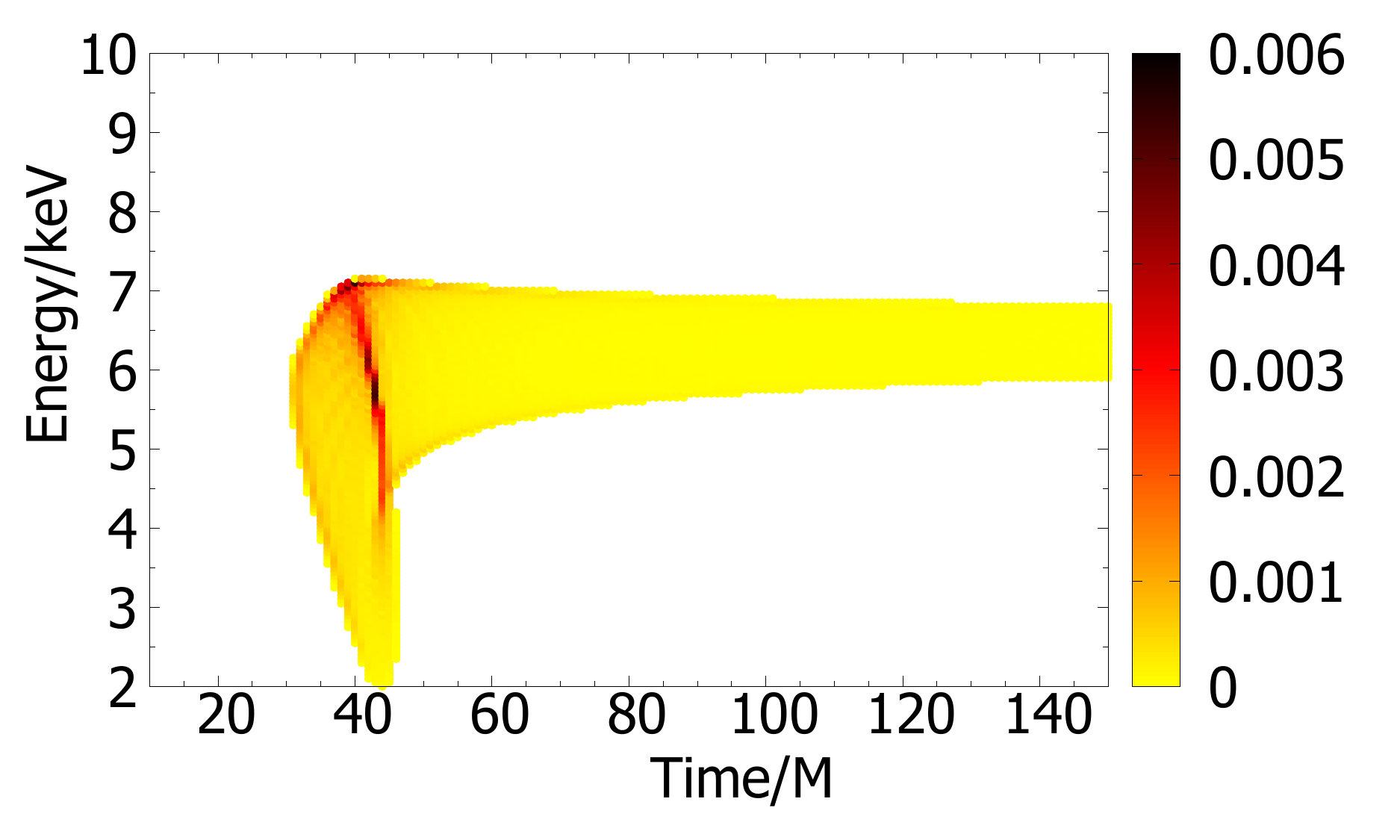}
\hspace{0.5cm}
\includegraphics[type=pdf,ext=.pdf,read=.pdf,width=7.0cm]{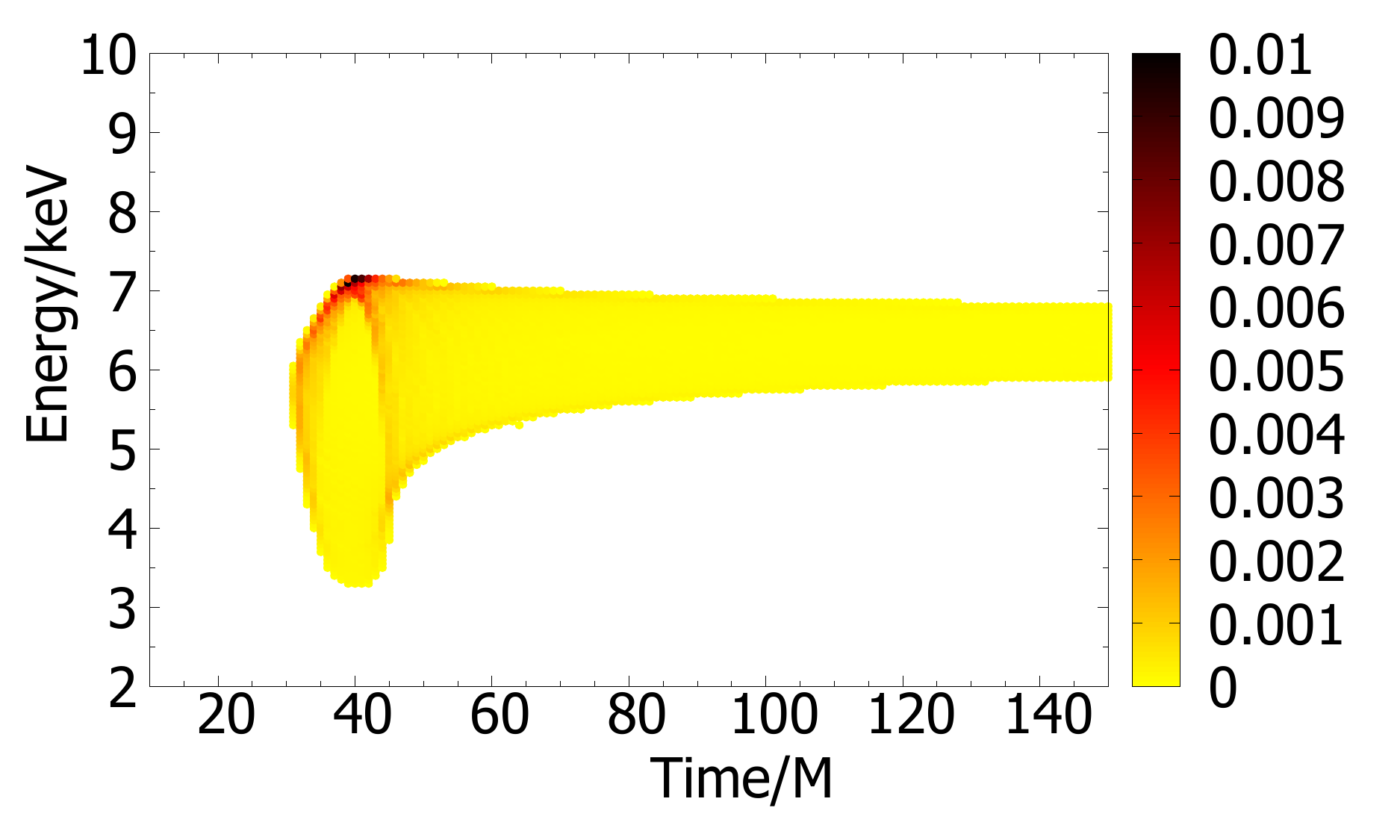}
\end{center}
\vspace{-0.5cm}
\caption{Transfer functions for Johannsen-Psaltis BHs with spin parameter
$a_*=0.5$ and deformation parameter $\epsilon_3 = 5$ (left panel) and
$\epsilon_3 = -5$ (right panel). The height of the source is $h = 10 \, M$ and
the index of the intensity profile is $q = 3$. 
Flux in arbitrary units. See the text for more details.}
\label{fig5}
\end{figure}

\section{The 2D transfer function for non-Kerr backgrounds \label{s-bh}}

In order to verify the Kerr nature of an astrophysical BH candidate, it is not enough
to observe relativistic signatures absent in a Newtonian model and have a good fit
to the BH's spectrum, because a non-Kerr BH may mimic a Kerr BH with a
different spin parameter. Such a degeneracy appears to be a common feature 
of non-Kerr structures.  In order to quantify possible deviations from the Kerr 
background, we have to consider a general BH metric specified by a mass $M$, 
spin parameter $a$, and additional free parameters (usually called deformation 
parameters) that must recover the Kerr solution as they vanish. The calculations 
of the theoretical predictions of the BH spectrum is performed within this more 
general spacetime, and the comparison between predictions and observations 
are used to measure the value of all the parameters of the background metric. 
The Kerr BH hypothesis is verified if observations agree with vanishing deformation 
parameters.

In principle, the test-metric should be sufficiently general that,
for arbitrary values of the deformation parameters, it would be possible to recover any BH
solution in any theory of gravity. However, at present there is no general framework 
that can take arbitrary deviations from the Kerr metric into
account. In addition, there is a strong correlation between the
measured spin parameter and deformation parameters measuring departure
from the Kerr solution.  Because of this, current observations are
severely challenged in assessing even one deformation
parameter. Therefore, we restrict our attention to the {\em simplest} scenario, in which the
object is allowed to be either more oblate or more prolate than a Kerr BH
with the same spin and its departure from the Kerr solution is regulated by a single 
deformation parameter. We accordingly adopt the simplest version of the 
Johannsen-Psaltis metric which employs one deformation parameter~\cite{jpm}. 
In Boyer-Lindquist coordinates, the line element reads 
\be\label{eq-jp}
ds^2 &=& - \left(1 - \frac{2 M r}{\Sigma}\right) \left(1 + h\right) dt^2
- \frac{4 a M r \sin^2\theta}{\Sigma} \left(1 + h\right) dtd\phi
+ \frac{\Sigma \left(1 + h\right)}{\Delta + a^2 h \sin^2 \theta} dr^2
+ \nonumber\\ &&
+ \Sigma d\theta^2 + \left[ \left(r^2 + a^2 +
\frac{2 a^2 M r \sin^2\theta}{\Sigma}\right) \sin^2\theta +
\frac{a^2 (\Sigma + 2 M r) \sin^4\theta}{\Sigma} h \right] d\phi^2 \, ,
\ee
where $\Sigma = r^2 + a^2 \cos^2\theta$, $\Delta = r^2 - 2 M r + a^2$, and
\be
h = \frac{\epsilon_3 M^3 r}{\Sigma^2} \, .
\ee
$\epsilon_3$ is the deformation parameter. The compact object is more prolate
(oblate) than a Kerr BH for $\epsilon_3 > 0$ ($\epsilon_3 < 0$); when $\epsilon_3 = 0$,
we exactly recover the Kerr solution.

The reverberation calculations for a Kerr background as treated in the
previous section can be quite naturally extended to 
Johannsen-Psaltis spacetime. Now, the parameters of the spacetime
geometry are $a_*$ and $\epsilon_3$ ($M$ simply sets the scale of the
system but otherwise factors out of any calculation).  Fig.~\ref{fig5}
shows the transfer functions for BHs with spin parameter $a_* = 0.5$
and non-vanishing $\epsilon_3$. They can be compared with the transfer
function in the top right panel in Fig.~\ref{fig1}, where the only
difference is that for Fig.~\ref{fig1} the deformation parameter is
zero; that is, the object is a Kerr BH. The most important difference
between these transfer functions is that $\epsilon_3$ affects the ISCO
radius and thus the minimum photon energy. For $\epsilon_3 = 5$ (left
panel in Fig.~\ref{fig5}), the inner edge of the disk is closer to the
BH than in the Kerr solution, so photons experience a stronger
gravitational redshift and the transfer function extends to energies
lower than 2~keV.  For $\epsilon_3 = - 5$ (right panel in
Fig.~\ref{fig5}), the object is more oblate than its Kerr counterpart,
so the gravitational force on the equatorial plane is stronger and the
ISCO radius is larger. In this spacetime, photons coming from the
inner part of the accretion disk are less affected by the
gravitational redshift as compared to cases with $\epsilon_3=5$ or
$\epsilon_3 = 0$, so the red wing of the 2D transfer function is less
extended. However, the transfer function is imprinted with more
information about the spacetime geometry. As a result, different
backgrounds experience different Doppler boosting, different light
bending, and different time delays. All these relativistic effects are
encoded in the 2D transfer function.

\begin{figure}
\begin{center}
\includegraphics[type=pdf,ext=.pdf,read=.pdf,width=7.0cm]{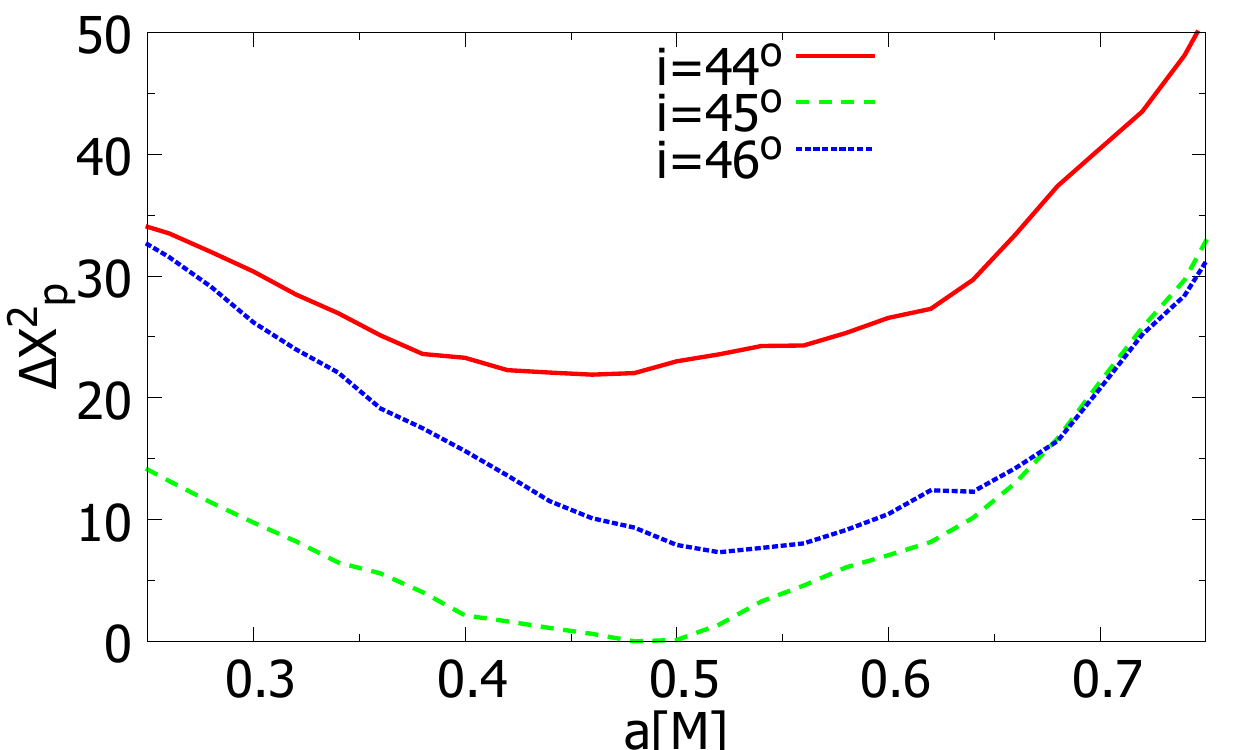}
\hspace{0.5cm}
\includegraphics[type=pdf,ext=.pdf,read=.pdf,width=7.0cm]{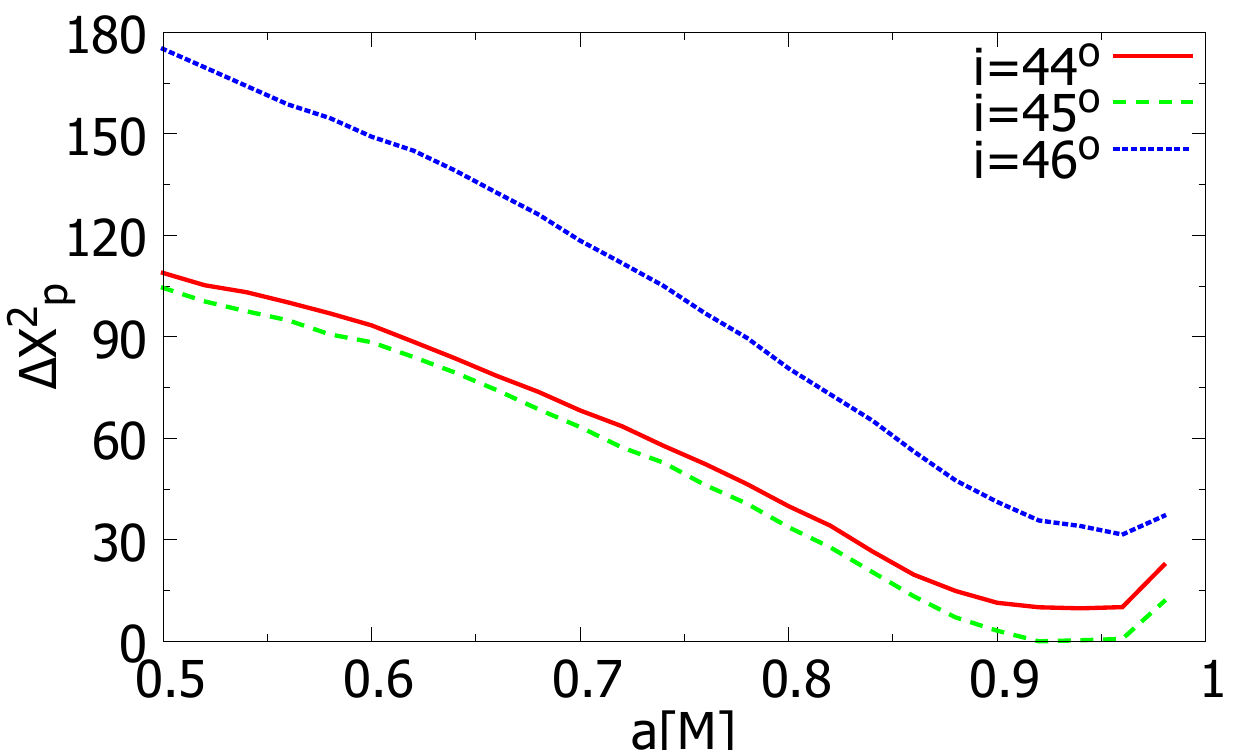}
\end{center}
\vspace{-0.5cm}
\caption{Left panel: 
The change in $\chi_{\rm p}^2$ compared to its minimum; 
$\chi_{\rm p}^2 \approx N \mathcal{L}_{\rm p}$ with $N=10^3$
from the comparison of the iron line profile of a Kerr BH
with spin parameter $a_*' = 0.5$ and inclination angle $i' = 45^\circ$
and the iron line profile of Kerr BHs with spin parameter $a_*$ and inclination
angle $i = 44^\circ$, $45^\circ$, and $46^\circ$. Right panel: as in the left panel
with a reference iron line profile of a Kerr BH with spin parameter $a_*' = 0.95$
and inclination angle $i' = 45^\circ$. The index of the intensity function is
always assumed to be $q = q' = 3$. 
See the text for more details.}
\label{fig6}
\vspace{0.6cm}
\begin{center}
\includegraphics[type=pdf,ext=.pdf,read=.pdf,width=7.0cm]{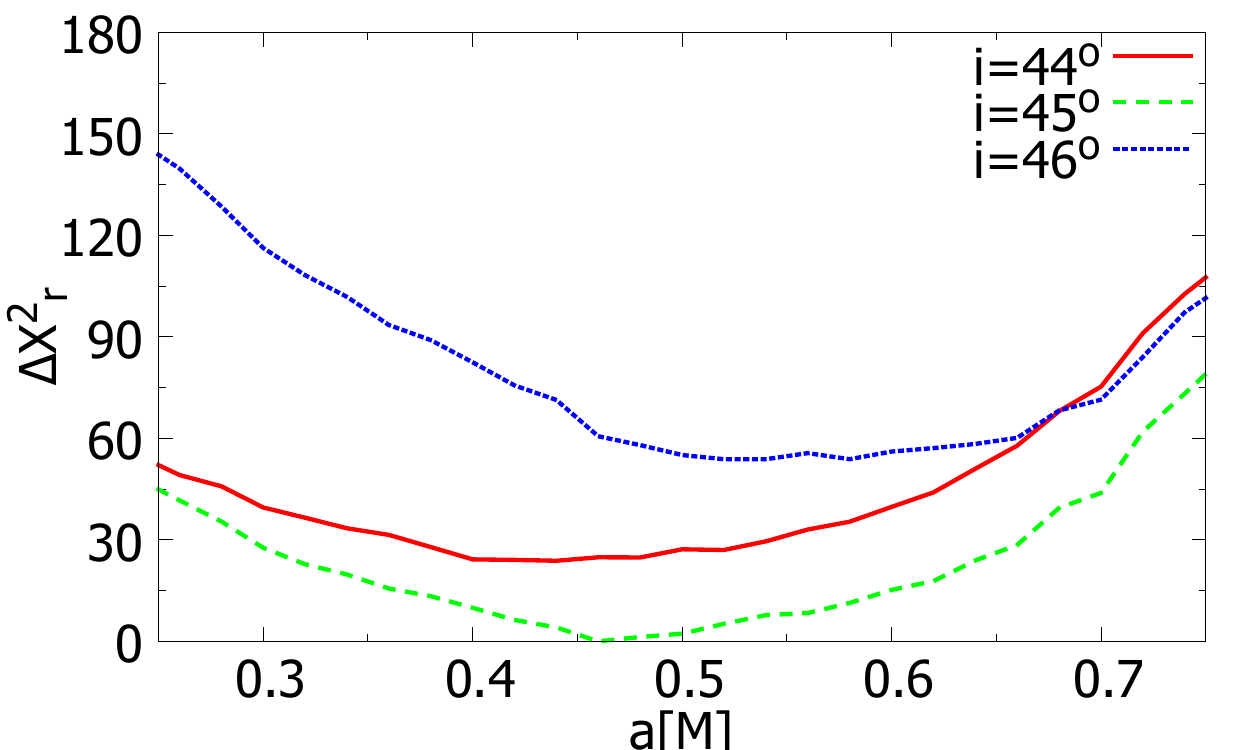}
\hspace{0.5cm}
\includegraphics[type=pdf,ext=.pdf,read=.pdf,width=7.0cm]{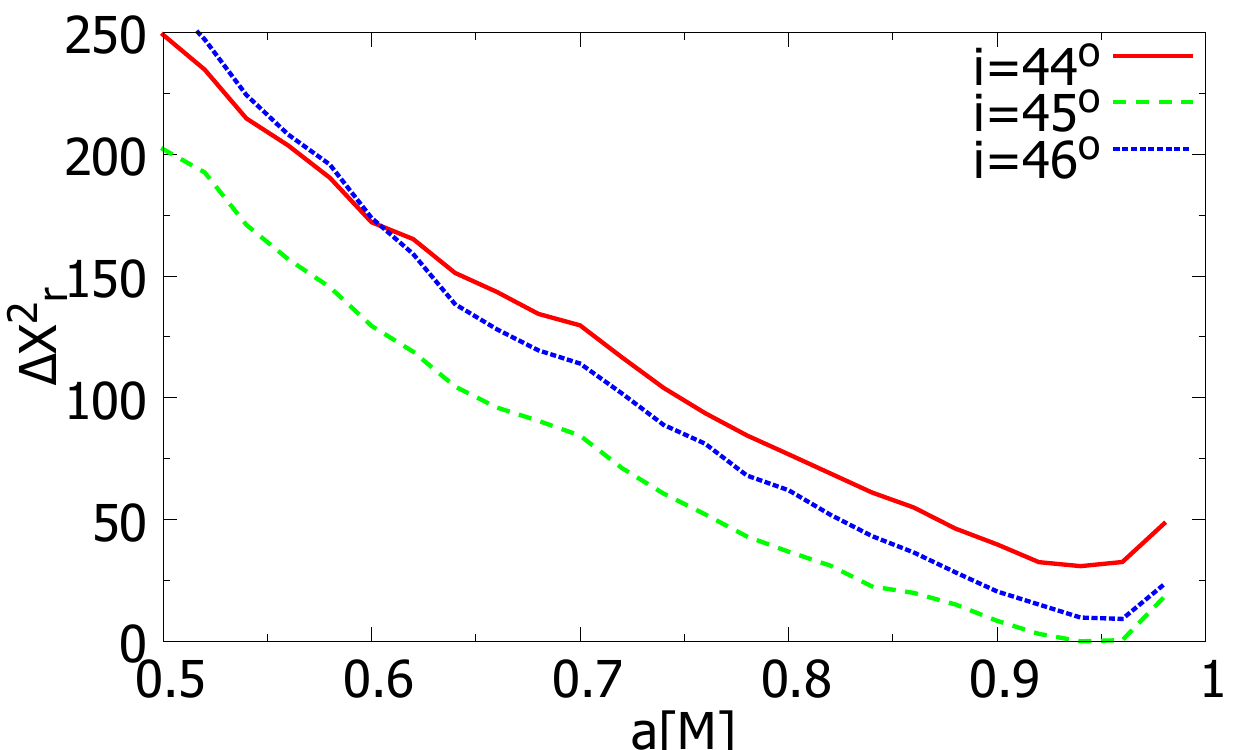}
\end{center}
\vspace{-0.5cm}
\caption{Left panel: 
Change in $\chi_{\rm r}^2$ compared to its minimum; 
$\chi_{\rm r}^2 \approx N \mathcal{L}_{\rm r}$ with $N=10^3$
from the comparison of the 2D transfer function of a 
Kerr BH with spin parameter $a_*' = 0.5$ and inclination angle $i' = 45^\circ$ and the 
2D transfer function of Kerr BHs with spin parameter $a_*$ and inclination angle 
$i = 44^\circ$, $45^\circ$, and $46^\circ$. 
The time range of the 2D transfer function is $[0, 300 \, M]$. 
Right panel: as in the left panel with a 
reference 2D transfer function of a Kerr BH with spin parameter $a_*' = 0.95$ and 
inclination angle $i' = 45^\circ$. The index of the intensity function is always assumed 
to be $q = q' = 3$ and the height of the source $h = h' = 10 \, M$. 
See the text for more details.}
\label{fig7}
\end{figure}

\begin{figure}
\begin{center}
\includegraphics[type=pdf,ext=.pdf,read=.pdf,width=7.0cm]{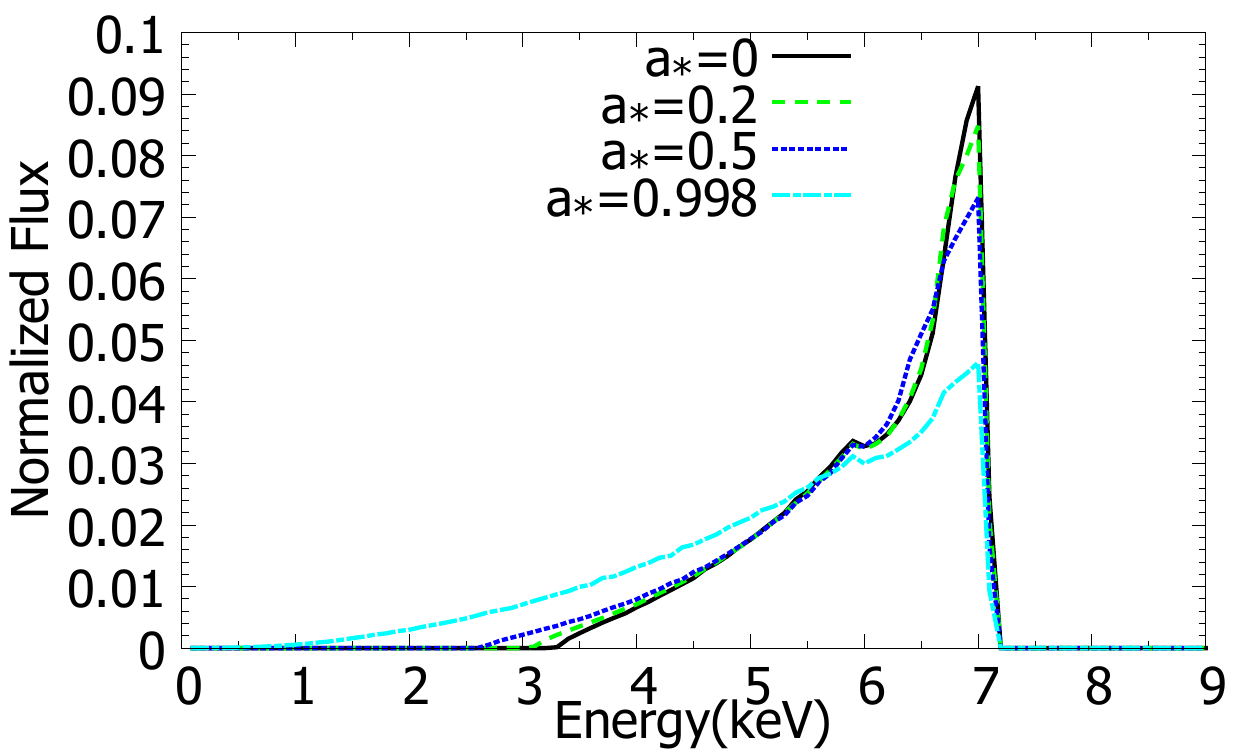}
\hspace{0.5cm}
\includegraphics[type=pdf,ext=.pdf,read=.pdf,width=7.0cm]{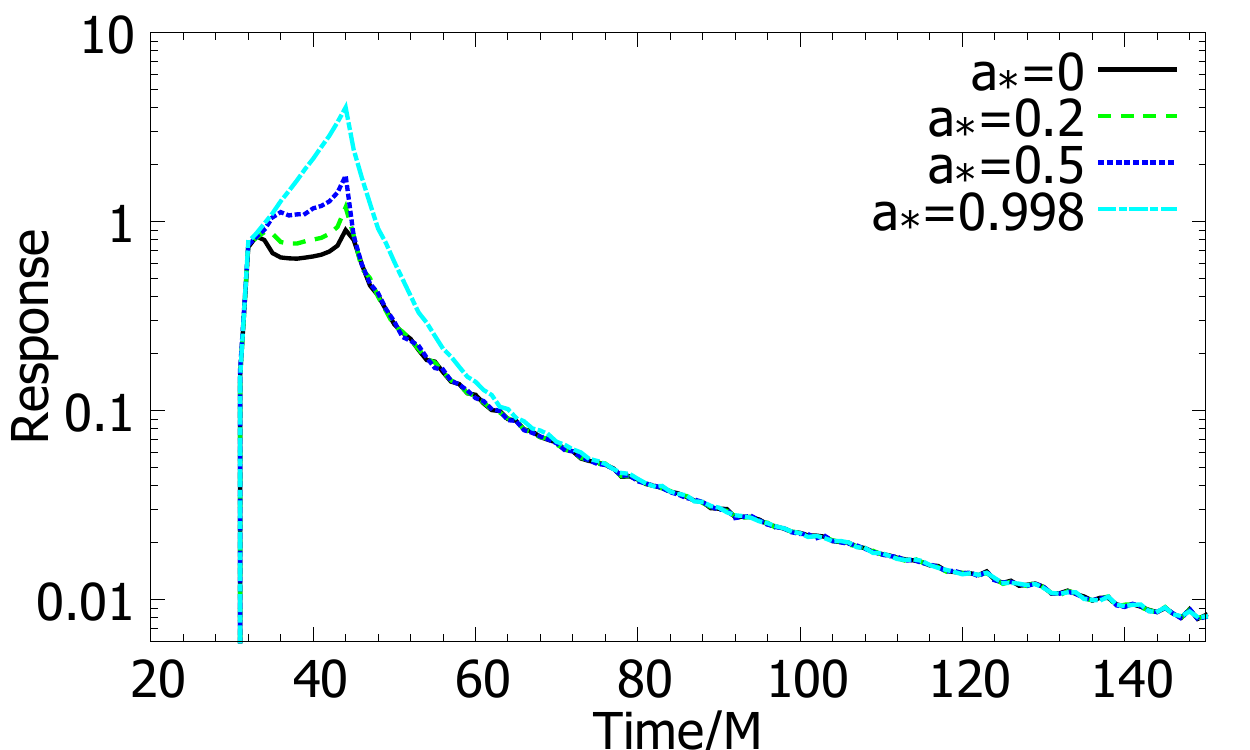}
\end{center}
\vspace{-0.5cm}
\caption{Left panel: iron line profile (time-integrated 2D 
transfer function) for Kerr BHs with spin $a_*=0$, 0.2, 0.5, and 0.998, an
inclination angle $i=45^\circ$, and an emissivity index $q=3$. 
Right panel: as in the left panel for the response function (energy-integrated
2D transfer function) and a source height $h = 10 \, M$. See the text for more details. }
\label{fig8}
\end{figure}

\begin{figure}
\begin{center}
\includegraphics[type=pdf,ext=.pdf,read=.pdf,width=7.0cm]{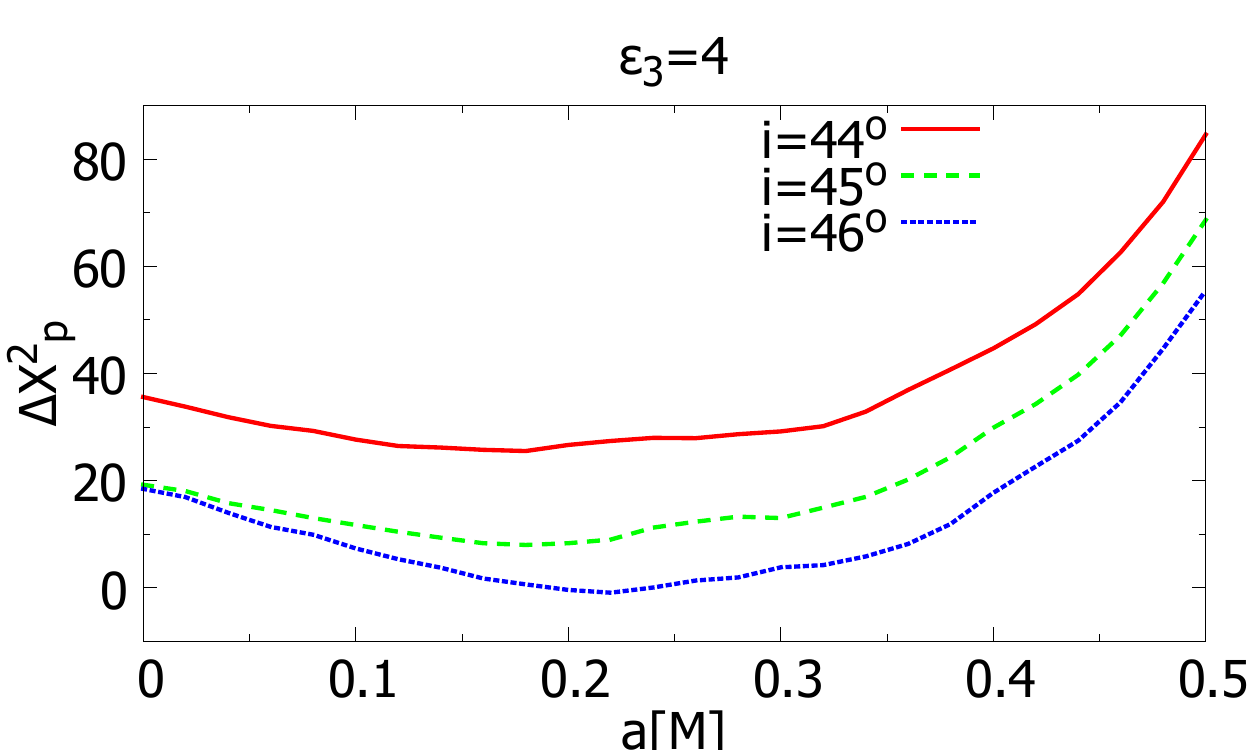}
\hspace{0.5cm}
\includegraphics[type=pdf,ext=.pdf,read=.pdf,width=7.0cm]{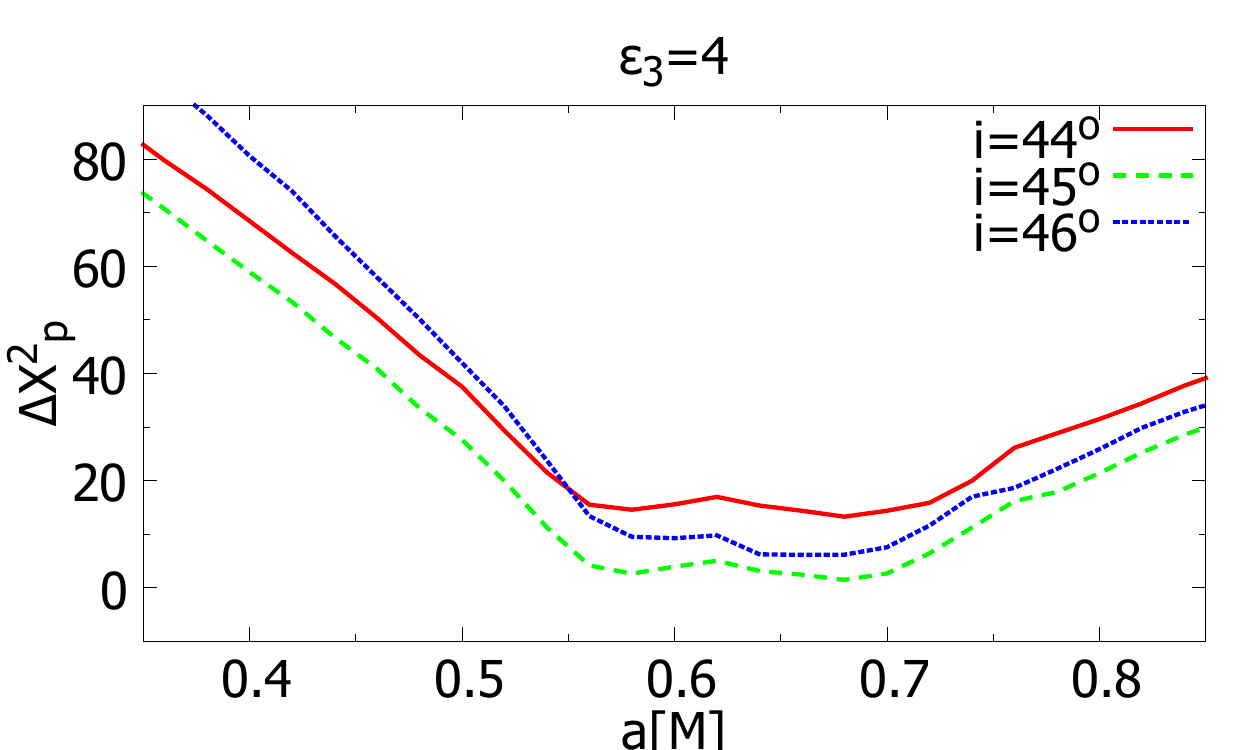}
\end{center}
\vspace{-0.5cm}
\caption{Left panel: 
Change in $\chi_{\rm p}^2$ compared to its minimum; 
$\chi_{\rm p}^2 \approx N \mathcal{L}_{\rm p}$ with $N=10^3$
from the comparison of the iron line profile of a Kerr 
BH with spin parameter $a_*' = 0.5$ and inclination angle $i' = 45^\circ$ and the iron 
line profile of Johannsen-Psaltis BHs with spin parameter $a_*$,
deformation parameter $\epsilon_3 = 4$,  
and inclination angle $i = 44^\circ$, $45^\circ$, and $46^\circ$. Right 
panel: as in the left panel with a reference iron line profile of a Kerr BH with spin 
parameter $a_*' = 0.95$. The index of the intensity function is always assumed to be 
$q = q' = 3$. See the text for more details.}
\label{fig9}
\vspace{0.6cm}
\begin{center}
\includegraphics[type=pdf,ext=.pdf,read=.pdf,width=7.0cm]{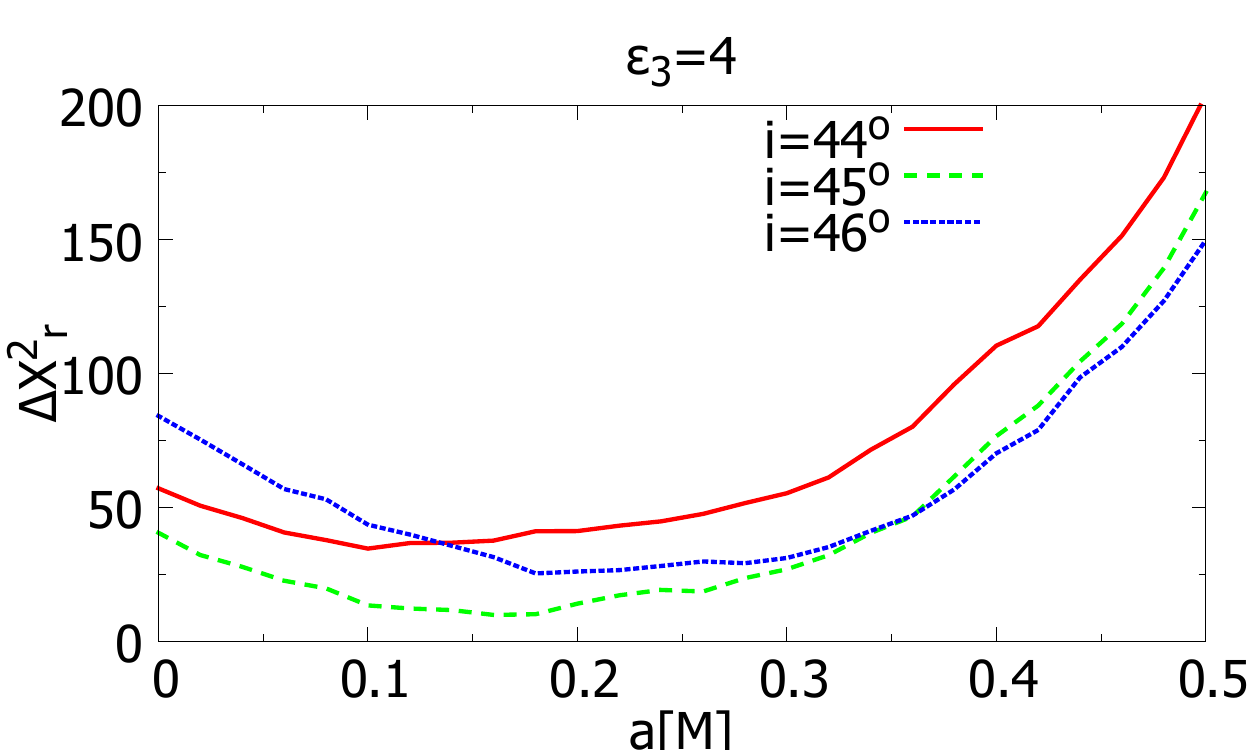}
\hspace{0.5cm}
\includegraphics[type=pdf,ext=.pdf,read=.pdf,width=7.0cm]{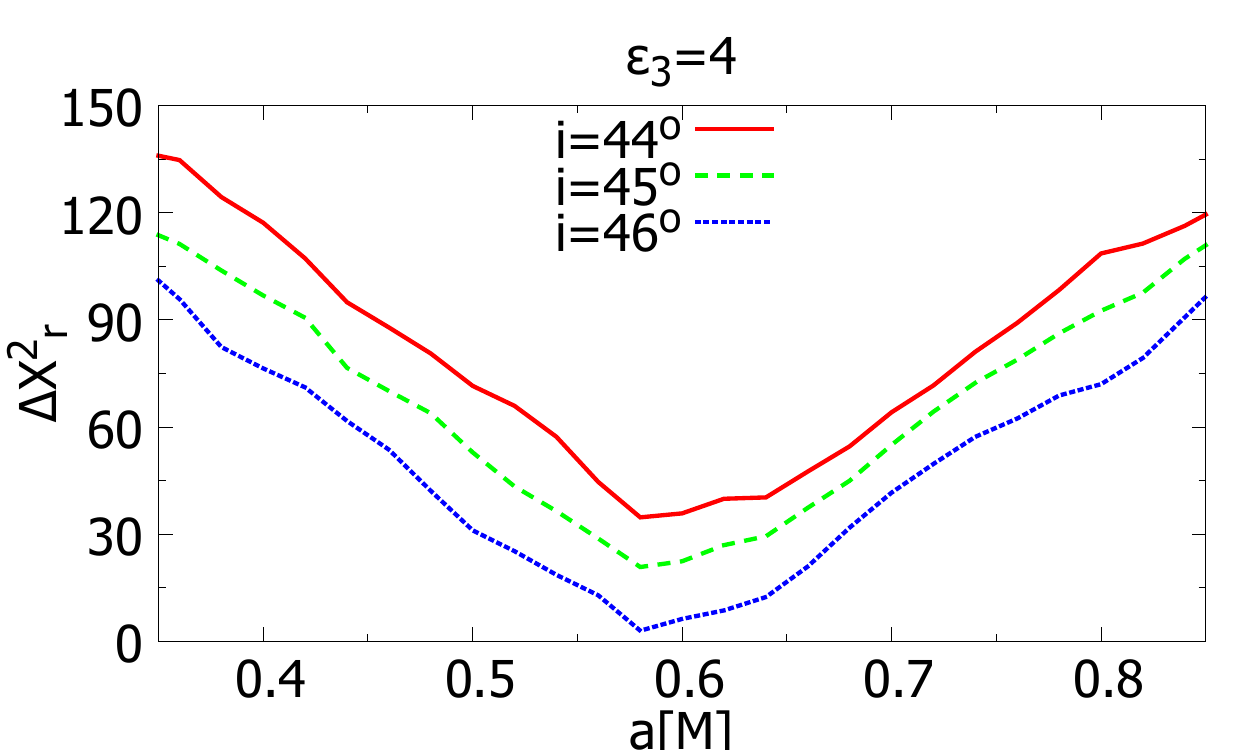}
\end{center}
\vspace{-0.5cm}
\caption{Left panel:
Change in $\chi_{\rm r}^2$ compared to its minimum; 
$\chi_{\rm r}^2 \approx N \mathcal{L}_{\rm r}$ with $N=10^3$
from the comparison of the 2D transfer function of a 
Kerr BH with spin parameter $a_*' = 0.5$ and inclination angle $i' = 45^\circ$ and the 
2D transfer function of Johannsen-Psaltis BHs with spin parameter $a_*$,
deformation parameter $\epsilon_3 = 4$, 
and inclination angle $i = 44^\circ$, $45^\circ$, and $46^\circ$. 
The time range of the 2D transfer function is $[0, 300 \, M]$. 
Right panel: as in the left panel with a reference 2D transfer function of a Kerr BH with spin 
parameter $a_*' = 0.95$. The index of the intensity function is always assumed to be 
$q = q' = 3$ and the height of the source $h = h' = 10 \, M$.
See the text for more details.}
\label{fig10}
\end{figure}

\begin{figure}
\begin{center}
\includegraphics[type=pdf,ext=.pdf,read=.pdf,width=7.0cm]{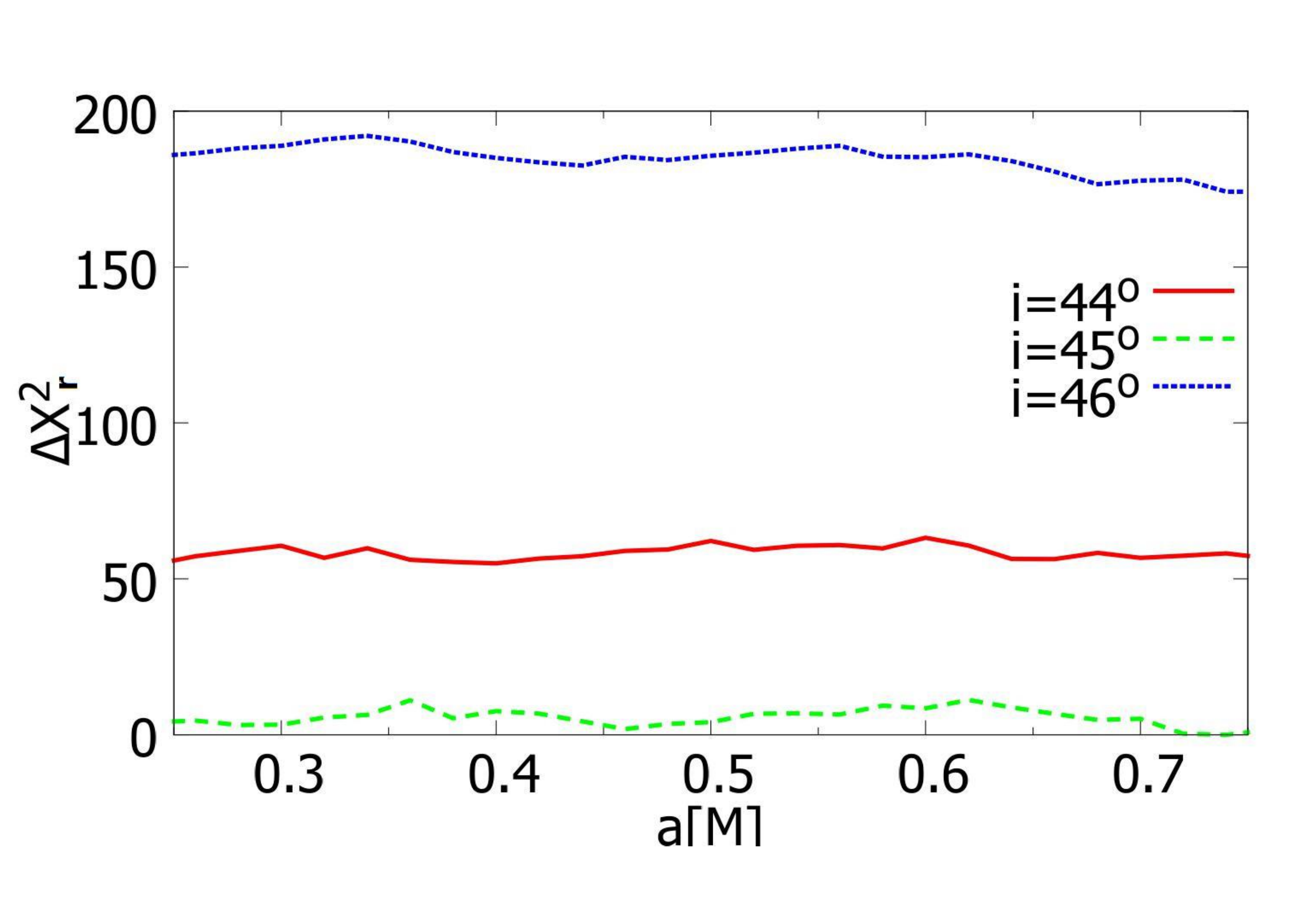}
\hspace{0.5cm}
\includegraphics[type=pdf,ext=.pdf,read=.pdf,width=7.0cm]{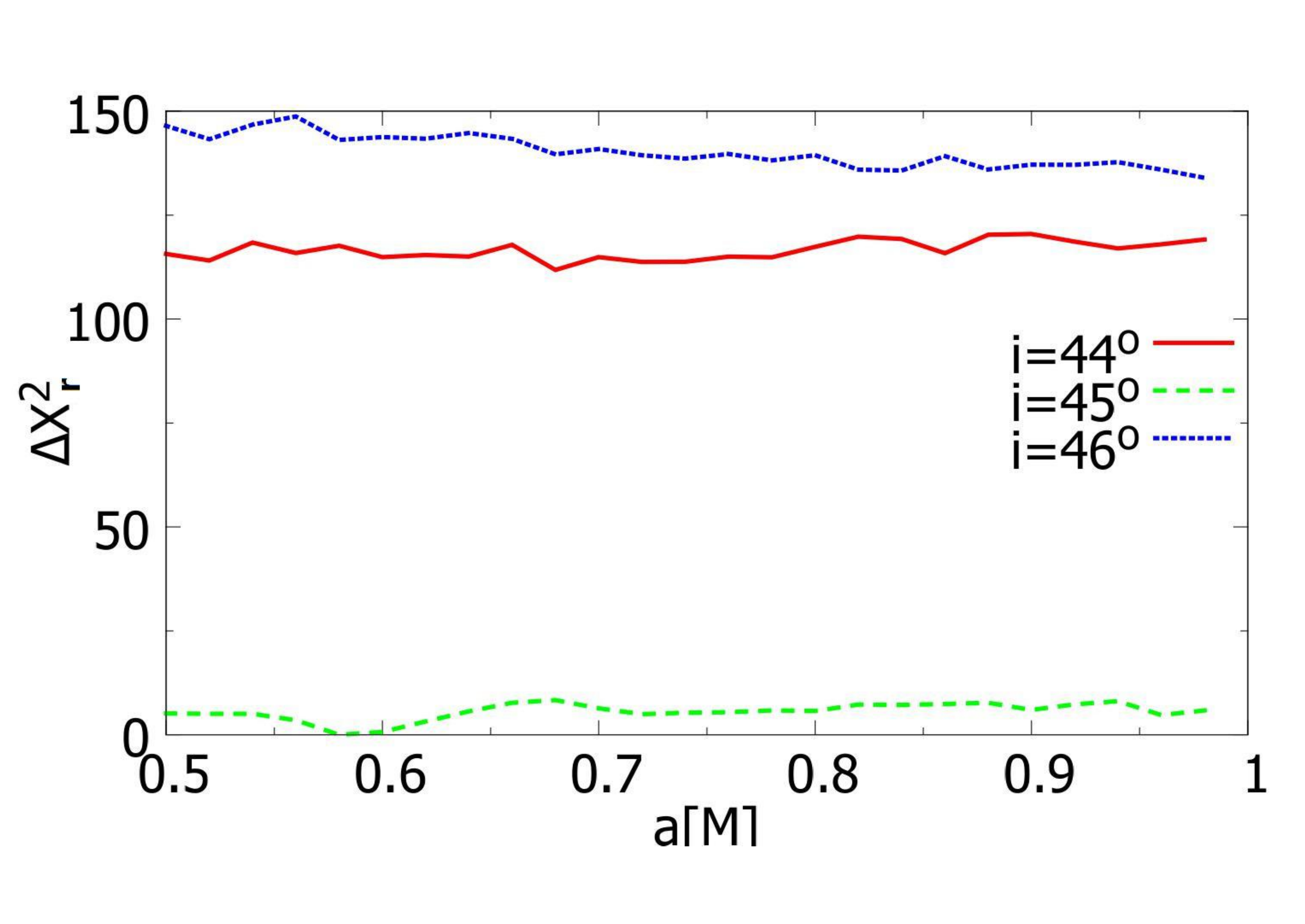}
\end{center}
\vspace{-0.5cm}
\caption{Left panel: 
As in Fig.~\ref{fig7}, but for the time range $[100\, M, 300\, M]$ 
which contains $N = 5000$. See the text for more details.}
\label{fig11}
\vspace{0.6cm}
\begin{center}
\includegraphics[type=pdf,ext=.pdf,read=.pdf,width=7.0cm]{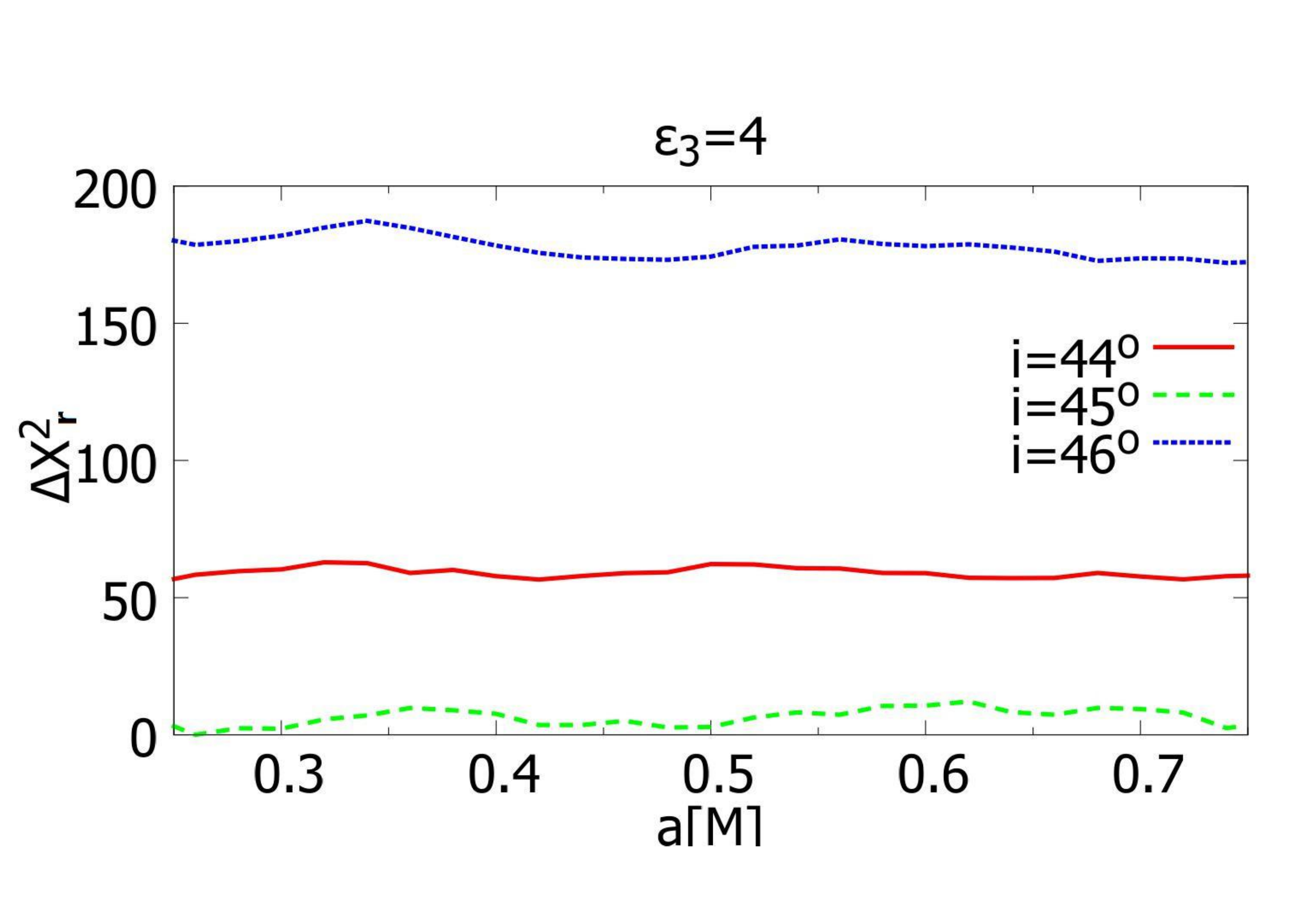}
\hspace{0.5cm}
\includegraphics[type=pdf,ext=.pdf,read=.pdf,width=7.0cm]{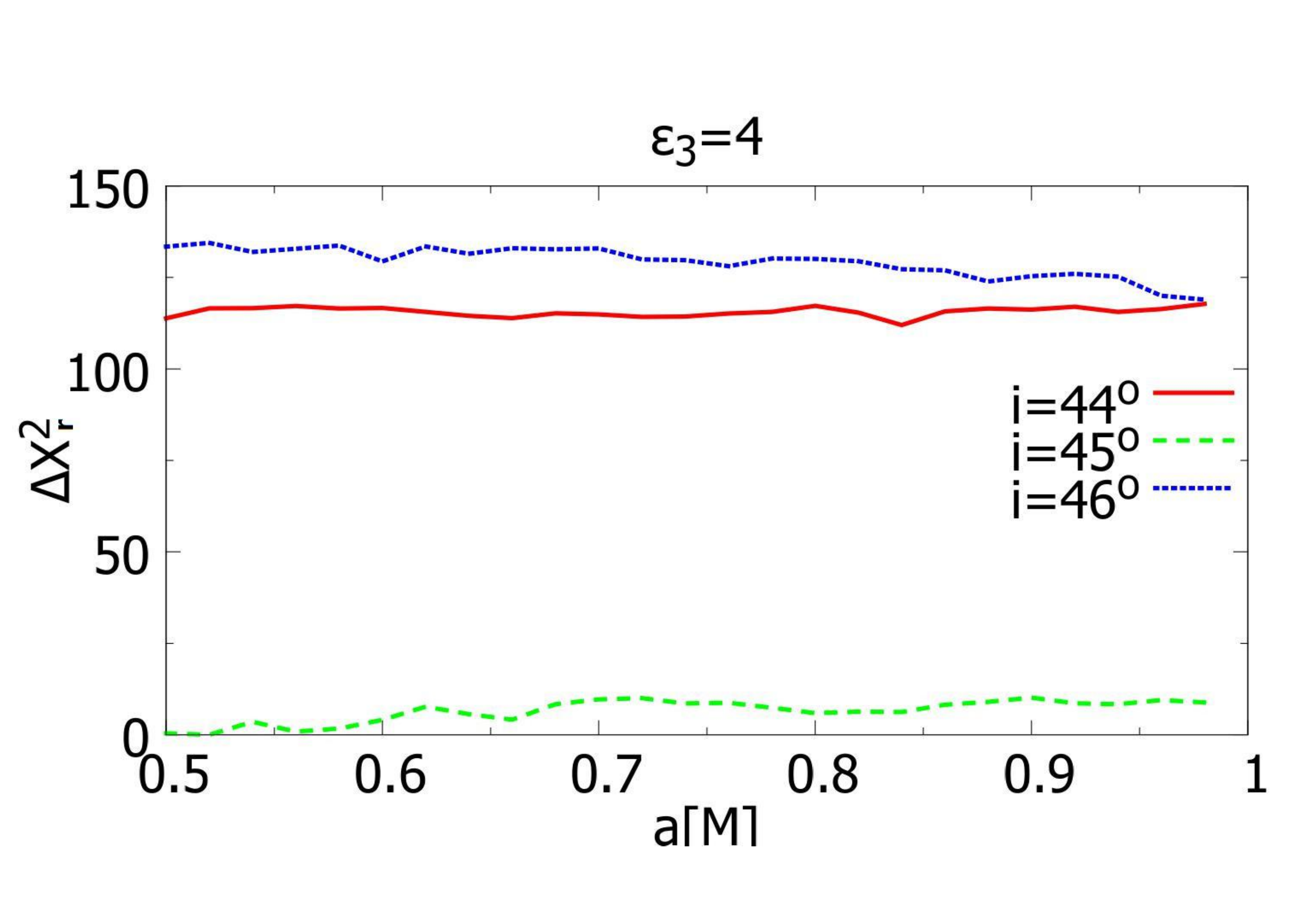}
\end{center}
\vspace{-0.5cm}
\caption{Left panel:
As in Fig.~\ref{fig10}, but the time range $[100\, M, 300\, M]$ 
which contains $N = 5000$. See the text for more details.} 
\label{fig12}
\end{figure}

\section{Comparison of Kerr and non-Kerr models \label{s-sim}}

Now we want to compare line profiles and reverberation transfer 
functions of Kerr and non-Kerr models. Let us consider
a primary model for which the BH has spin parameter $a_*$, deformation 
parameter $\epsilon_3$, the 
disk is observed from an inclination angle $i$, the emissivity index is $q$ and 
the height of the on-axis X-ray source is $h$.  
Here we employ a single inclination angle of the disk, namely we do not consider the possibility that the disk is warped at these scales.
We use the notation
\be 
n_j = n(a_*, \epsilon_3,i,q)
\ee
to indicate the photon flux number density in the energy bin 
$[E_j, E_j + \Delta E]$ of the iron line profile.  Similarly, for the
2D transfer function, we employ the notation
\be 
n_{jk} = n(a_*, \epsilon_3,i,q,h)
\ee
for the photon flux number density in the energy bin $[E_j, E_j + \Delta E]$ 
and in the time bin $[t_k, t_k + \Delta t_k]$.  We take a secondary model to be
compared with the primary for which the BH has spin parameter $a_*'$, 
deformation parameter $\epsilon_3'$, 
the disk is observed from an inclination angle $i'$, the emissivity index is $q'$ and 
the height of the on-axis X-ray source is $h'$.   The secondary model
line profile and transfer function intensities are expressed,
respectively, as
$n_j' = n(a_*', \epsilon_3',i',q')$ and $n_{jk}' = n(a_*', \epsilon_3',i',q',h')$. 
We  introduce the normalized (negative) log-likelihood $\mathcal{L}_{\rm p}$ 
and $\mathcal{L}_{\rm r}$ for means of comparison between the primed and 
unprimed iron line {\em profiles} and 2D {\em reverberation} transfer functions.  We adopt notations of ${\rm p}$ and ${\rm r}$ to emphasize which situation is being considered: 
\be \label{log-l1}
\mathcal{L}_{\rm p} &=& 
\frac{1}{\sum_{j} n_{j}} \left[
\sum_{j} \frac{\left(n_{j} - \alpha_{\rm p} n_{j}'\right)^2}{n_{j}} \right] \, , \\
\mathcal{L}_{\rm r} &=& \frac{1}{\sum_{j,k} n_{jk}} \left[
\sum_{j,k} \frac{\left(n_{jk} - \alpha_{\rm r} n_{jk}'\right)^2}{n_{jk}}
\right] \, , \label{log-l2}
\ee 
where $\alpha_{\rm p,r}$ is chosen to minimize $\mathcal{L}_{\rm p,r}$, namely
\be
\alpha_{\rm p} =  \frac{\sum_{j} n_{j}'}{\sum_{j} n_{j}'^2/n_{j}} \, , \quad
\alpha_{\rm r} = \frac{\sum_{j,k} n_{jk}'}{\sum_{j,k} n_{jk}'^2/n_{jk}} \, .
\ee
The corresponding chi-square is, respectively, 
$\chi_{\rm p}^2 \approx N \mathcal{L}_{\rm p}$ and $\chi_{\rm r}^2 \approx N \mathcal{L}_{\rm r}$, 
where $N$ is the number of detected photons. 
Given a mission sensitivity,
we can estimate which features can potentially be distinguished 
and at which level of confidence. 
In what follows, we use either
$N = 10^3$, corresponding to a high-quality observation today,
and $N = 10^5$, which is an optimistic benchmark for a quality observation with a next generation of X-ray satellite.
Moreover, we adopt
$\Delta E = 0.05$~keV and $\Delta t = M$ (e.g., for $M = 10^6 \, M_\odot$, 
 $\Delta t \approx 5$~s). 
$M$ is also a model parameter in the transfer function, but its value 
would often be obtained externally using optical data (e.g.,~\cite{petebook}).
Moreover, in a fit it should be correlated with the height of the source, 
in the sense that the height and mass are degenerate for a given lag~\cite{cackett}, 
but not with the parameters of the spacetime geometry close to the BH.

We simulate observations of iron lines from our models for an arbitrary $N$ photons, incorporating Poisson noise, and treat the simulation as if it were data.  To this end, our analysis employs standard techniques, consisting of grouping the data to a minimum of 10 counts per bin.   Because of these observation-mimicking procedures, the relationship between $\chi^2$ and $N \mathcal{L}$ is only approximate (it becomes exact in the limit of high counts).
Here we adopt a minimum of 10 counts per bin which achieves a 3-sigma significance per bin.  We note that this choice is arbitrary; there are other reasonable choices of binning; however, this is of minor consequence and has no impact on our results.

\subsection{Role of the model parameters}

In this Paper, our emphasis is on correlation between the 
spin measurement, deformation parameter, and the viewing angle.  
Although other parameter correlations may also be explored, we find
that the spin and viewing angle are more correlated with the
deformation parameter 
measurement than e.g., the emissivity index or the height of the source.
While in this {\em exploratory} work we maintain a more narrow focus, 
we note that followup work seeking precise spin or
deformation parameter determination should fully examine degeneracy 
between all model parameters.

In Fig.~\ref{fig6}, we compare the {\em time-integrated} iron line profiles of Kerr BHs 
with spin parameter $a_*' = 0.5$ (left panel) and $a_*' = 0.95$ (right panel) observed 
with an inclination angle $i' = 45^\circ$ and the iron line profiles of Kerr BHs with
spin parameter $a_*$ ($x$ axis) and observed with an inclination angle $i = 44^\circ$
(red solid line), $45^\circ$ (green dashed line), and $46^\circ$ (blue dotted
line). The emissivity index is always assumed to be $q = q' = 3$. 
$\chi^2_{\rm p}$ is calculated with $N = 10^3$.
In Fig.~\ref{fig7}, we 
show $\chi_{\rm r}^2$, the chi-square obtained from the comparison 
of the 2D transfer functions. 
Here the time interval is $[0,300\;M]$ and $N$ is still $10^3$.
Figs.~\ref{fig6} and \ref{fig7} suggest that reverberation mapping can provide stronger constraints on the spacetime geometry, the improvement is significant but still modest, i.e., most of the information is already in the iron line profile. This has elsewhere been established for a pure Kerr model in Ref.~\cite{cackett}.
Fig.~\ref{fig8} shows the iron line profile (time-integrated 2D transfer function, left panel)
and the response function (energy-integrated 2D transfer function, right panel) of BHs
with different spin parameters.

In the same spirit, we can compare iron line profiles and 2D transfer functions of 
Kerr and Johannsen-Psaltis BHs. This is done, respectively, in Fig.~\ref{fig9} and 
Fig.~\ref{fig10}. In these plots, the reference spectra are again Kerr BHs with $a_*' = 0.5$ 
(left panels) and $a_*' = 0.95$ (right panels), with the same inclination angle 
$i' = 45^\circ$, emissivity index $q' = 3$, and height of the source $h = 10 \, M$, 
but they are compared to objects in the Johannsen-Psaltis metric with $\epsilon_3 = 4$. 
In these simulations, we have assumed $N=10^3$. The minimum 
of $\chi_{\rm p}^2$ is not always for $i = i'$. This is, in part, because there is a
correlation between the spin, the deformation parameter, and the inclination angle; 
that is, a non-Kerr object observed from a certain inclination angle may be interpreted 
as a Kerr BH with a different spin parameter and observed from a different inclination 
angle. This is exacerbated by the ``low'' photon count $N = 10^3$.

The advantages of the 2D transfer function is that we can potentially 
measure  effects from different radii, which is 
impossible using the time-averaged signal. However, a higher signal is necessary to employ this technique.
As one clear benefit, reverberation mapping can accurately pick out the inclination angle of 
the disk from the photons detected at later time, coming from larger radii. 
The photons detected in the time interval $[100 \, M, 300 \, M]$ are about 5\%
of the total photons in $[0, 300 \, M]$. For $N = 10^3$, the mere $\sim 50$
photons are insufficient signal.  For $N=10^5$, the yield is about 5000 photons
which provide a constraint on the inclination 
angle which is independent of the background geometry, as shown in Figs.~\ref{fig11} and \ref{fig12}. 
We note that the curves demonstrate how this late-time signal contains essentially no information on the spin.
As we will show in the next subsection, even the
time-integrated iron line profile with $N=10^5$ can correctly select the right
inclination angle. However, the determination of the inclination angle from late 
time photons can be considered robust, as it is independent of the 
reflection properties at small radii.

\begin{figure}
\begin{center}
\includegraphics[type=pdf,ext=.pdf,read=.pdf,width=7.0cm]{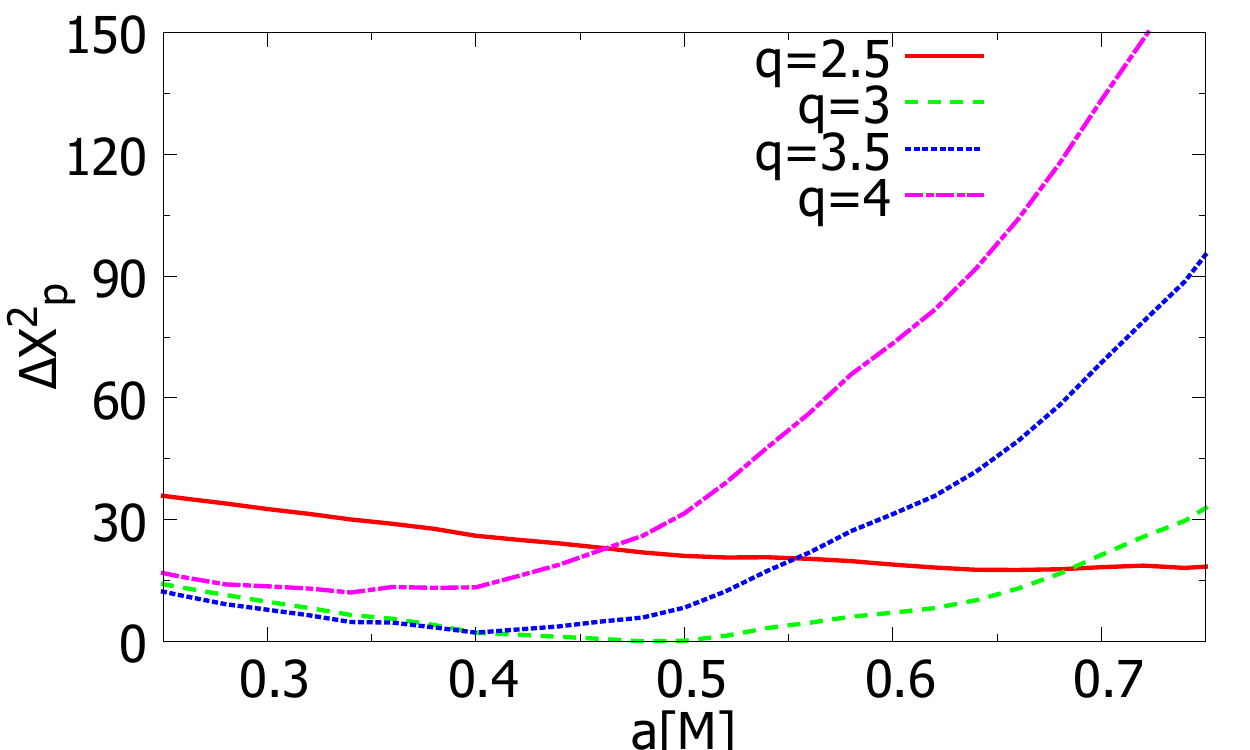}
\hspace{0.5cm}
\includegraphics[type=pdf,ext=.pdf,read=.pdf,width=7.0cm]{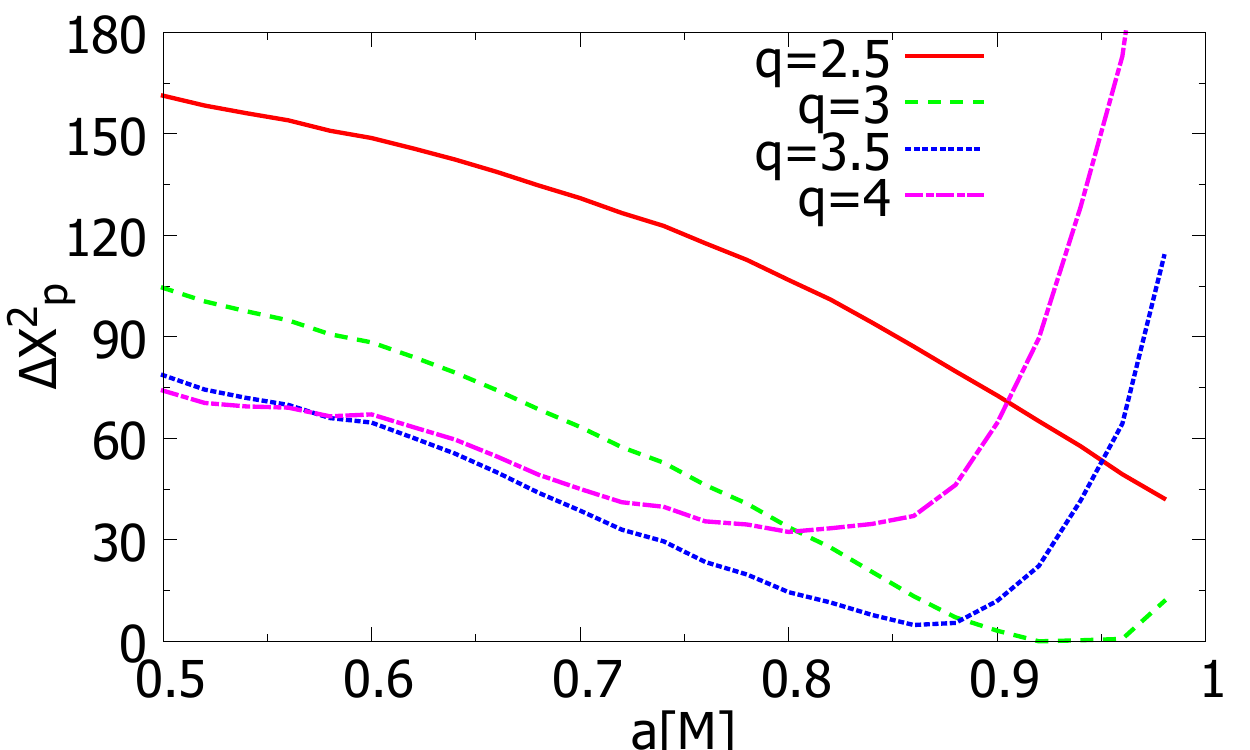}
\end{center}
\vspace{-0.5cm}
\caption{Left panel:
Change in $\chi_{\rm p}^2$ compared to its minimum; 
$\chi_{\rm p}^2 \approx N \mathcal{L}_{\rm p}$ with $N=10^3$
from a comparison of the iron line profile 
of a Kerr BH with spin parameter $a_*' = 0.5$ and emissivity index $q' = 3$ 
versus a set of Kerr BHs with emissivity indexes $q = 2.5$, 3, 3.5, and
4, as a function of spin $a_*$. Right panel: as in the left panel but with a reference Kerr 
BH with spin parameter $a_*' = 0.95$.  The inclination angle has been fixed to 
$45^\circ$ for all models. See the text for more details.}
\label{fig13}
\vspace{0.6cm}
\begin{center}
\includegraphics[type=pdf,ext=.pdf,read=.pdf,width=7.0cm]{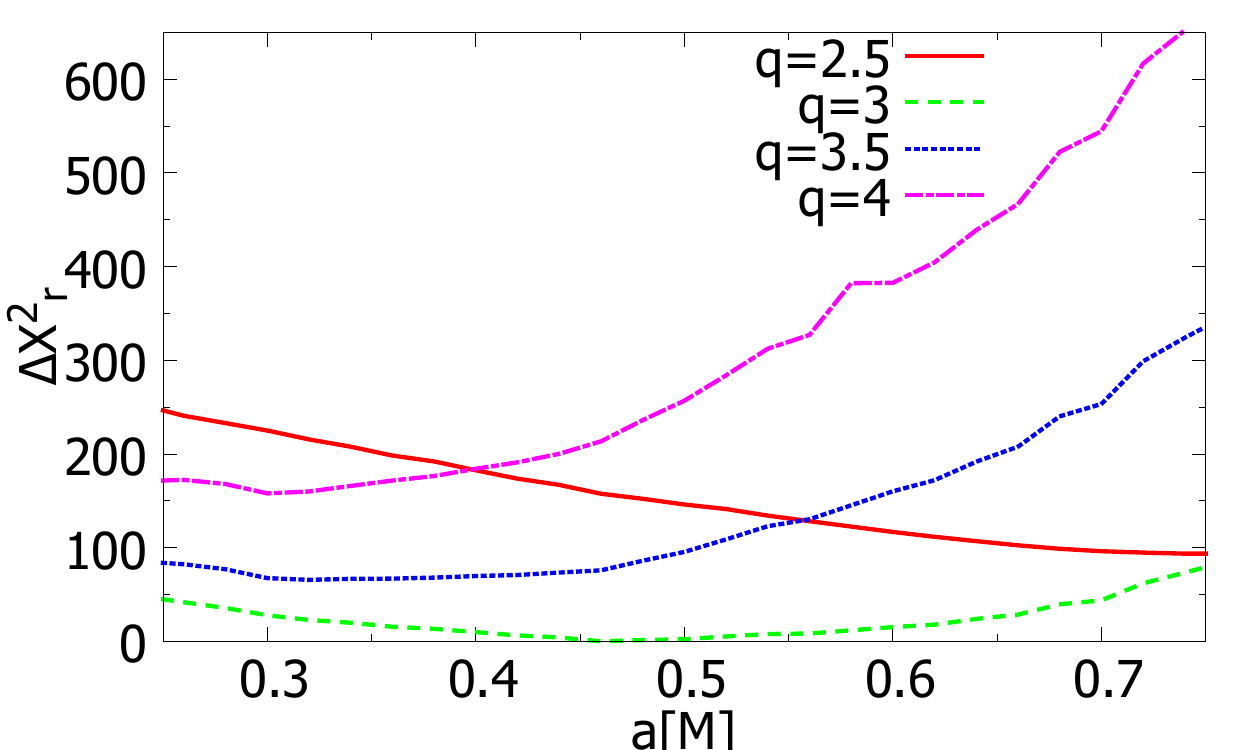}
\hspace{0.5cm}
\includegraphics[type=pdf,ext=.pdf,read=.pdf,width=7.0cm]{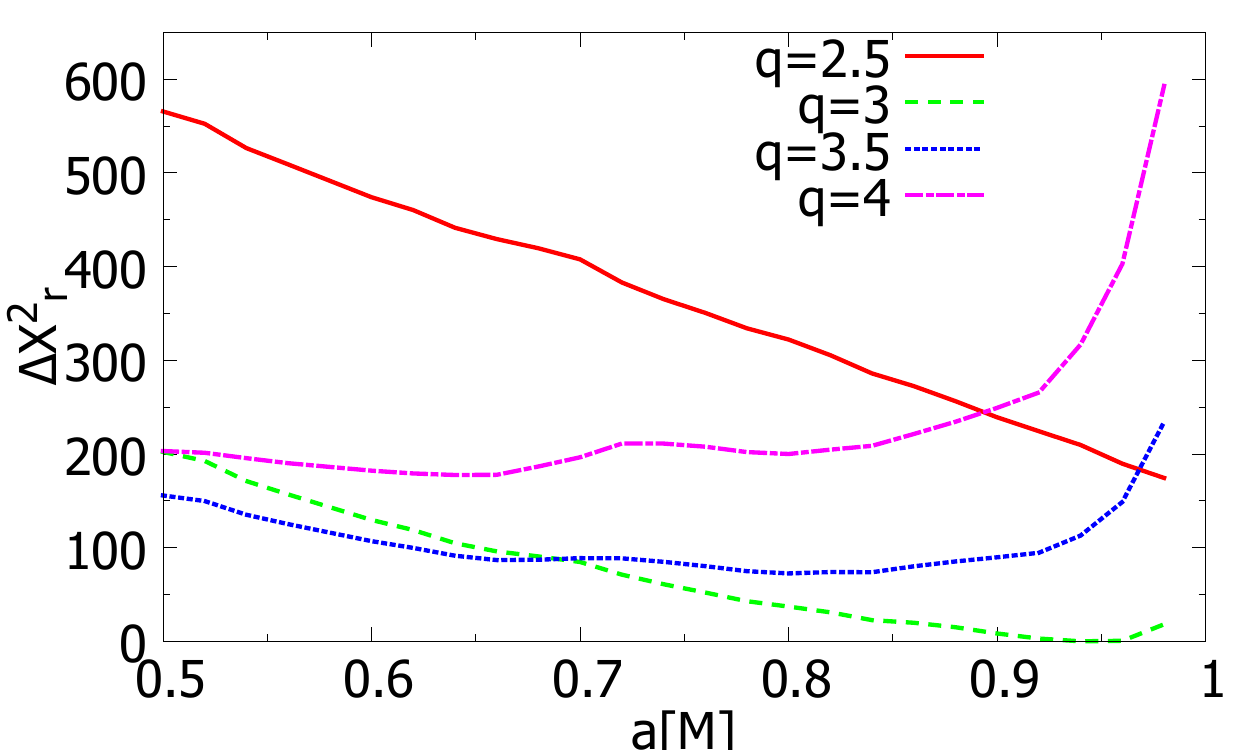}
\end{center}
\vspace{-0.5cm}
\caption{Left panel:
Change in $\chi_{\rm r}^2$ compared to its minimum; 
$\chi_{\rm r}^2 \approx N \mathcal{L}_{\rm r}$ with $N=10^3$
from a comparison of the 2D transfer 
function of a Kerr BH with spin parameter $a_*' = 0.5$ and emissivity index 
$q' = 3$  versus a set of Kerr BHs with emissivity indexes $q = 2.5$,
3, 3.5, and 4, as a function of spin $a_*$. 
The time range of the 2D transfer function is $[0, 300 \, M]$. 
Right panel: as in the left panel but with a 
reference Kerr BH with spin parameter $a_*' = 0.95$.  The
inclination angle has been fixed to $45^\circ$ and height of the
source taken as $10 \, M$ for all models. See the text for more details.}
\label{fig14}
\end{figure}

\begin{figure}
\begin{center}
\includegraphics[type=pdf,ext=.pdf,read=.pdf,width=7.0cm]{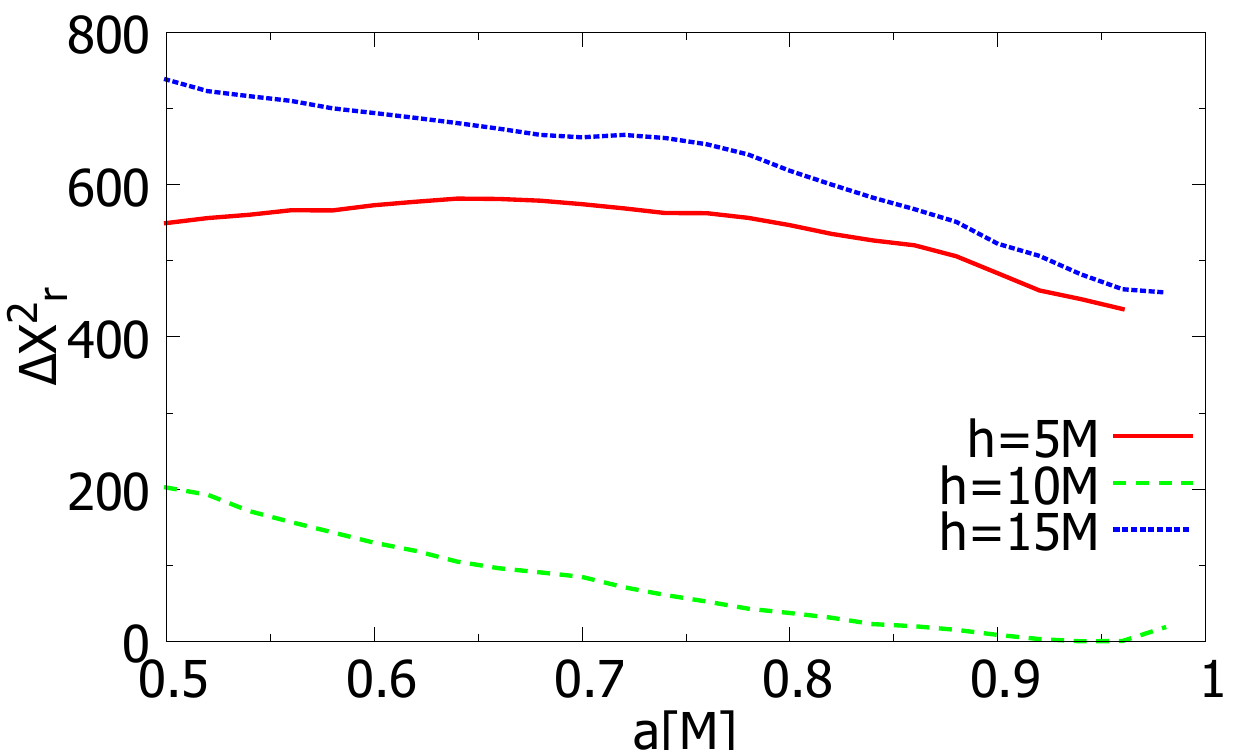}
\hspace{0.5cm}
\includegraphics[type=pdf,ext=.pdf,read=.pdf,width=7.0cm]{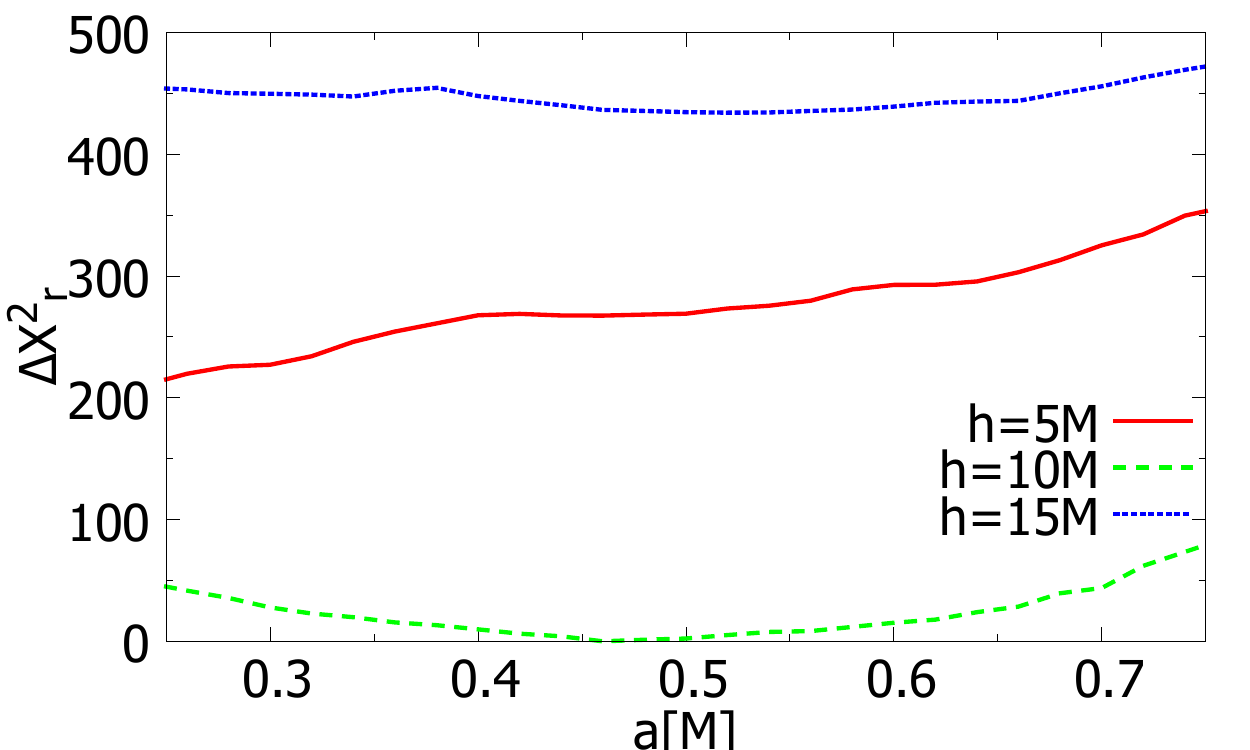}
\end{center}
\vspace{-0.5cm}
\caption{Left panel:
Change in $\chi_{\rm r}^2$ compared to its minimum; 
$\chi_{\rm r}^2 \approx N \mathcal{L}_{\rm r}$ with $N=10^3$
from a comparison of the 2D transfer 
function of a Kerr BH with spin parameter $a_*' = 0.5$ and a lamppost corona height 
$h' = 10 \, M$ versus a set of Kerr BHs with a range of corona height
$h = 5 \, M$, $10 \, M$, and $15 \, M$, as a function of spin $a_*$. 
The time range of the 2D transfer function is $[0, 300 \, M]$. 
Right panel: as in the left 
panel but with a reference Kerr BH with spin parameter 
$a_*' = 0.95$. The inclination angle is $i = i' = 45^\circ$ and emissivity index 
is $q = q' = 3$. See the text for more details.}
\label{fig15}
\end{figure}

The impact of the emissivity index on the iron line profile and the 2D
transfer function is shown, respectively, in Fig.~\ref{fig13} and Fig.~\ref{fig14}. For 
the sake of simplicity, we here fix the inclination angle and the height of the source.
It is important to note that one can explicitly link the height $h$ to $q(r)$, which requires more detailed modeling, e.g., \cite{dauser}.  However, this is beyond the scope of this exploratory work, in which we treat $q$ as a constant for simplicity.
The 2D transfer function can far better distinguish the effects of the
spin from those of the emissivity index with respect to the time-integrated spectrum: 
the curves for which emissivity index differs from the reference model are
well above the curve with the correct emissivity index across a wide spin range.  
By contrast, this is not the case for the time-integrated spectra, for which models
with erroneous emissivity indexes but spins close to the reference
value produce superior fits than the models with correct emissivity indexes but vastly different spins.
As shown in Ref.~\cite{iron3}, unlike inclination angle, the 
emissivity index constraint is very weakly correlated with the estimate of the deformation 
parameter from the iron line profile. Such a result can be qualitatively 
understood by noting that the deformation parameter affects the
redward extent of the 
low-energy tail (which relates to the inner edge of the disk and the relativistic effects 
there) as well as the position of the high-energy peak (which is produced by the 
Doppler boosting at larger radii, where the gravitational redshift is milder 
but the angular frequency of the orbit still high).  By contrast, the emissivity index 
alters the relative flux received by each region of the disk. It cannot move to lower or higher 
energies the position of the low-energy tail or of the most blueshifted photons.
In light of this and given the superior capability of reverberation mapping 
to distinguish the effects of 
spin from emissivity index, we argue that the correlation between the measurements 
of the deformation parameter and of the emissivity index from the 2D transfer function is
quite weak and can be reasonably neglected compared to the sizable 
correlation with the viewing angle $i$.

Lastly, we consider the role of the source height in
Fig.~\ref{fig15}.  In this case we cannot compare the iron line
profile with reverberation mapping because, in the prescription  
adopted, the former contains no information about this parameter.
As discussed in~\cite{cackett}, the measurement of the height of the source 
and the measurement of the mass are degenerate for a given lag,
in cases for which $M$ is not independently known. However, because $M$
is often externally constrained and given also that there is no
significant correlation between $M$ with the spacetime geometry
close to the putative BH, we keep $M$ fixed in our analysis for
purposes of examining possible tests of the Kerr metric. A detailed
consideration of the effect of keeping $M$ free is left to other work.
From Fig.~\ref{fig15} it seems that reverberation mapping is quite sensitive to the exact value of the height of the source. The models with $h = 5$~$M$ and 15~$M$ have a $\Delta \chi^2$ significantly larger than any model with the correct height. Such a constraining power can be presumably quite useful in models in which the index emissivity $q(r)$ depends on the height $h$~\cite{dauser}.

\begin{figure}
\begin{center}
\includegraphics[type=pdf,ext=.pdf,read=.pdf,width=7.2cm]{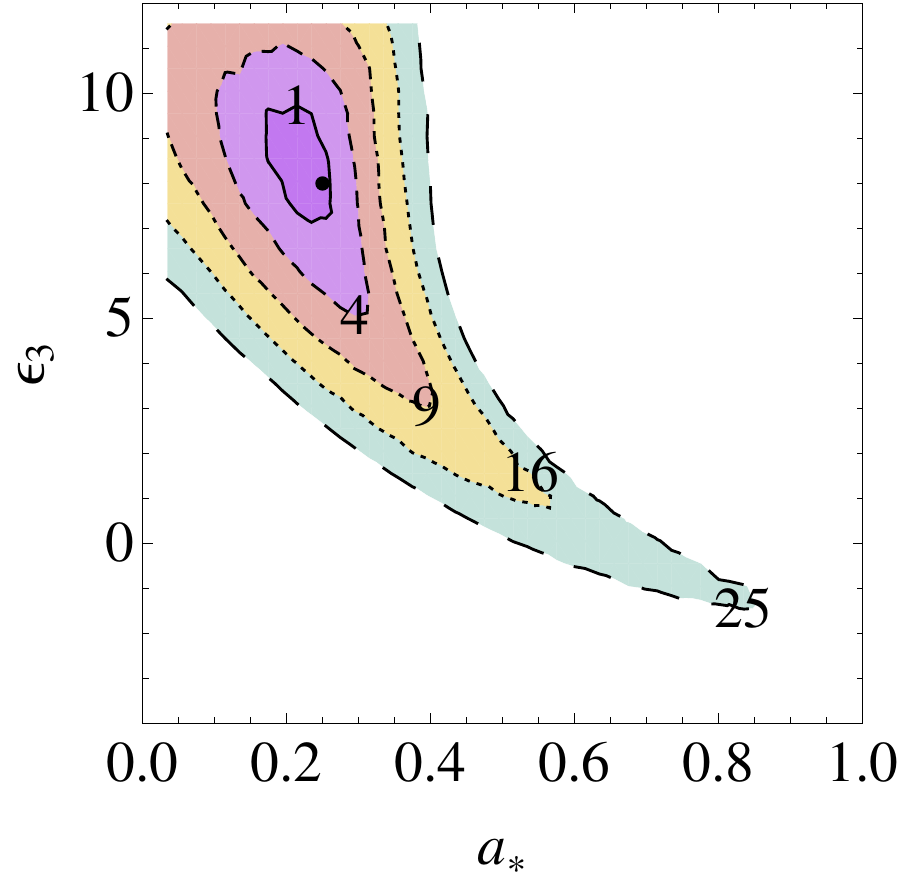}
\hspace{0.5cm}
\includegraphics[type=pdf,ext=.pdf,read=.pdf,width=7.2cm]{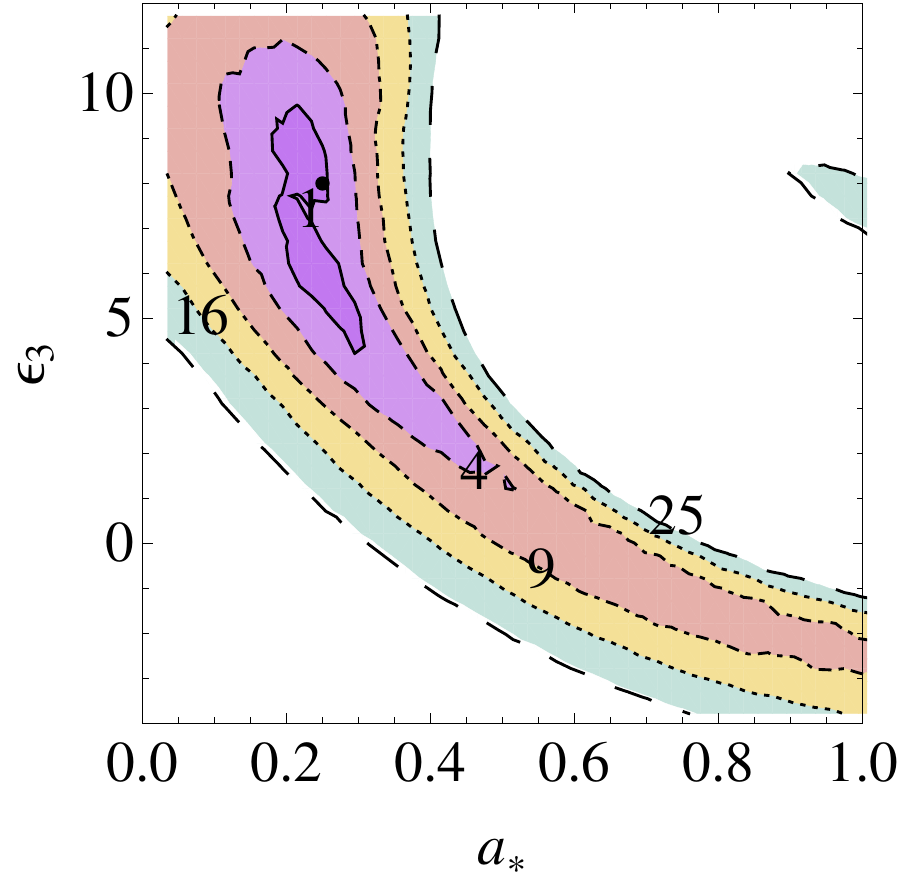} \\
\vspace{0.8cm}
\includegraphics[type=pdf,ext=.pdf,read=.pdf,width=7.2cm]{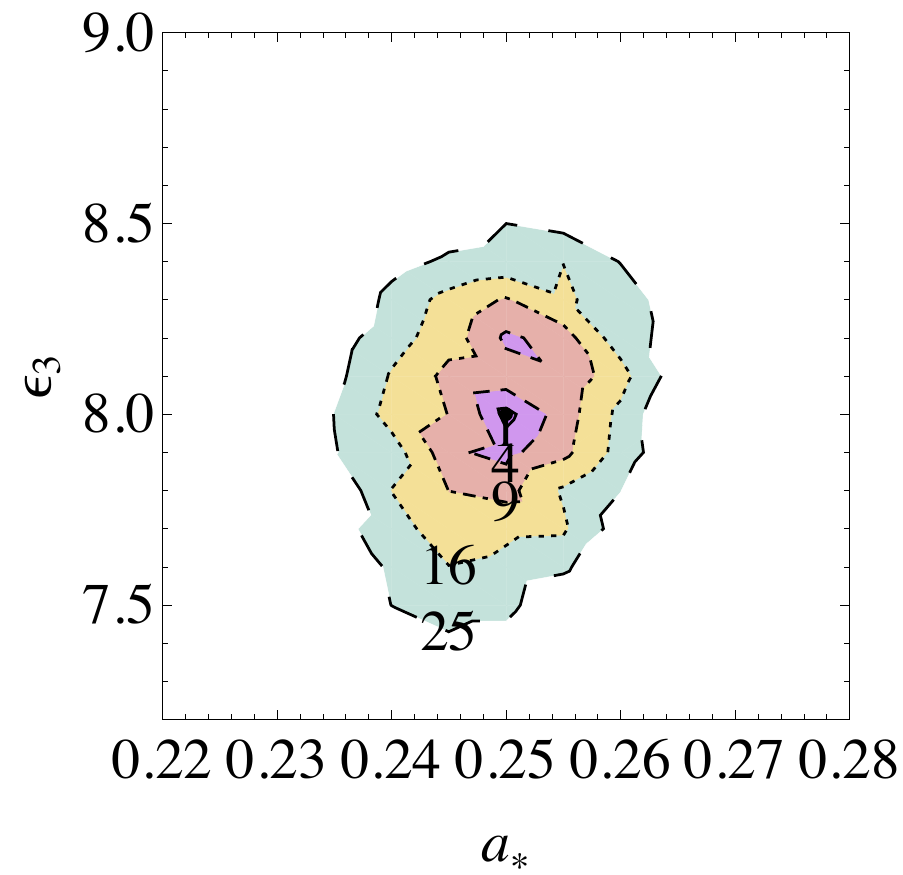}
\hspace{0.5cm}
\includegraphics[type=pdf,ext=.pdf,read=.pdf,width=7.2cm]{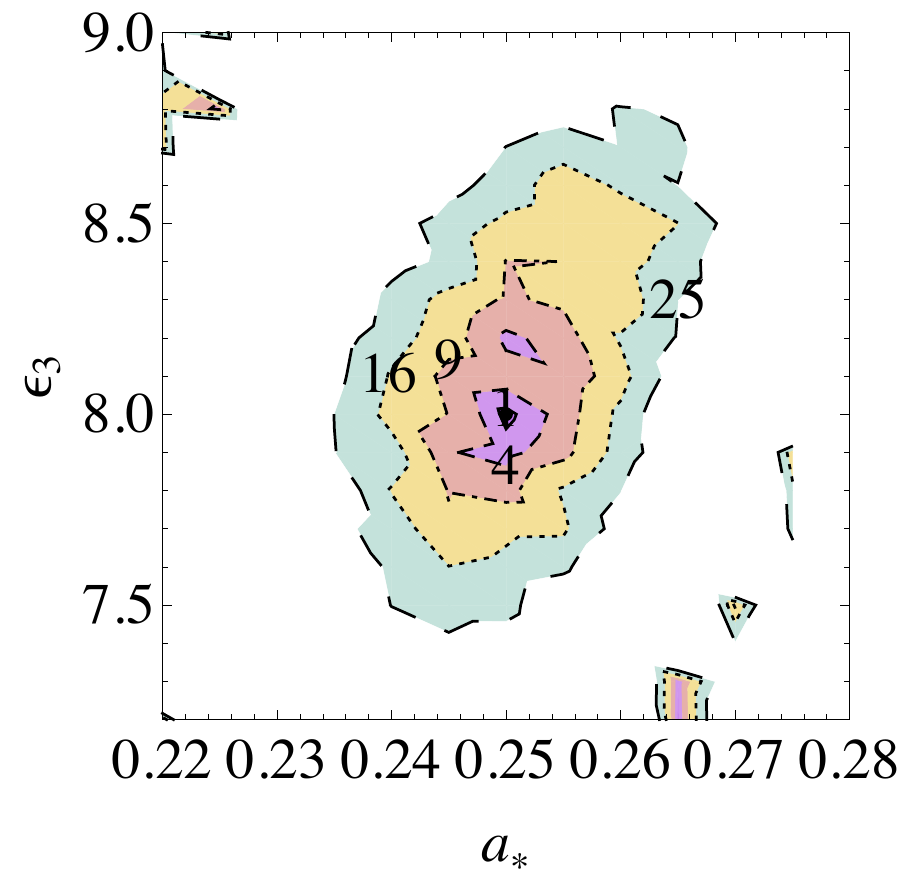}
\end{center}
\vspace{-0.5cm}
\caption{Top panels: $\Delta \chi_{\rm p}^2$ contours with $N=10^3$ from the comparison of the iron line profile 
of a Johannsen-Psaltis BH with a spin parameter $a_*' = 0.25$, deformation 
parameter $\epsilon_3' = 8$, and an inclination of $i'=45^\circ$ versus a set of 
Johannsen-Psaltis BHs with spin parameters $a_*$ ($x$ axis) and deformation 
parameters $\epsilon_3$ ($y$ axis). In the left panel, all models use an inclination 
angle $i'= i = 45^\circ$. In the right panel, $i$ is free in the fit. For the sake of simplicity, the emissivity index is 
fixed to $q' = q = 3$. Bottom panels: as in the top panels with $i = 45^\circ$ (left 
panel) and $i$ free (right panel) for $N=10^5$. See the text for more details.}
\label{fig16}
\end{figure}

\begin{figure}
\begin{center}
\includegraphics[type=pdf,ext=.pdf,read=.pdf,width=7.2cm]{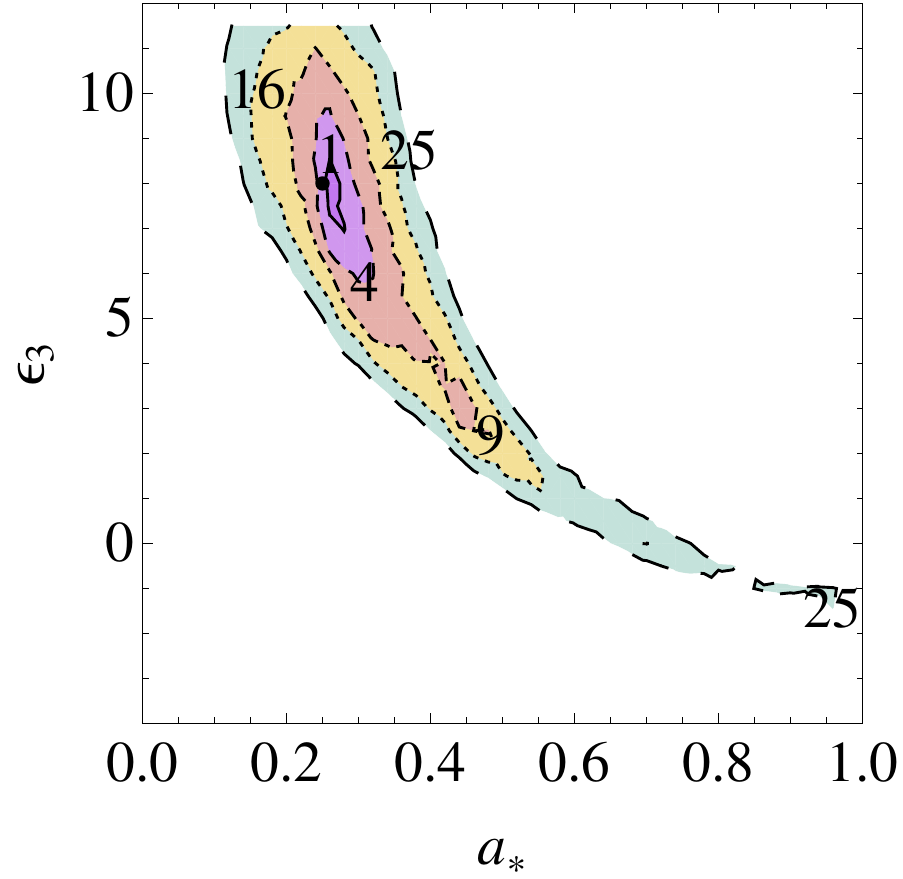}
\hspace{0.5cm}
\includegraphics[type=pdf,ext=.pdf,read=.pdf,width=7.2cm]{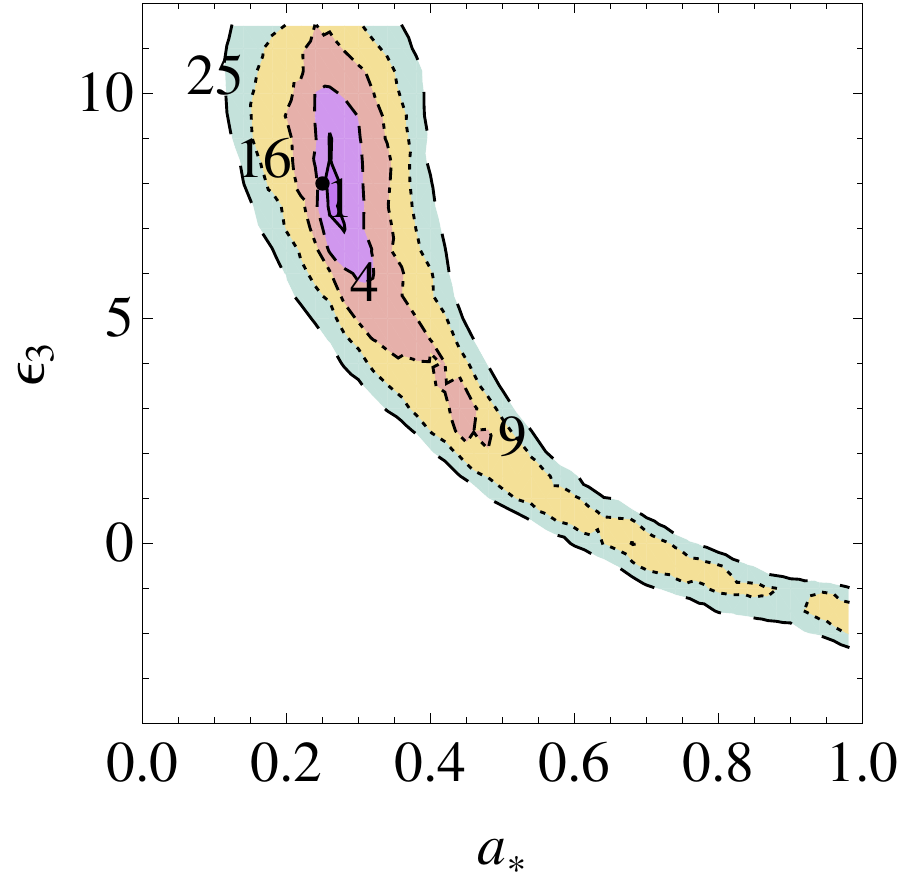} \\
\vspace{0.8cm}
\includegraphics[type=pdf,ext=.pdf,read=.pdf,width=7.2cm]{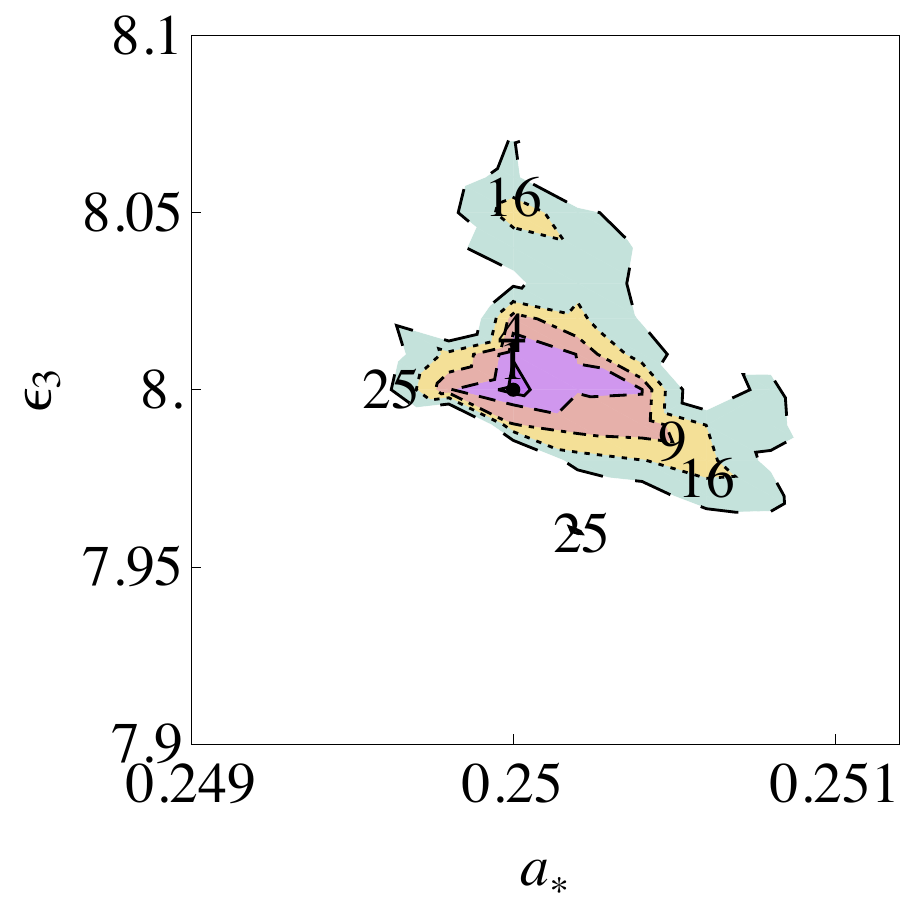}
\hspace{0.5cm}
\includegraphics[type=pdf,ext=.pdf,read=.pdf,width=7.2cm]{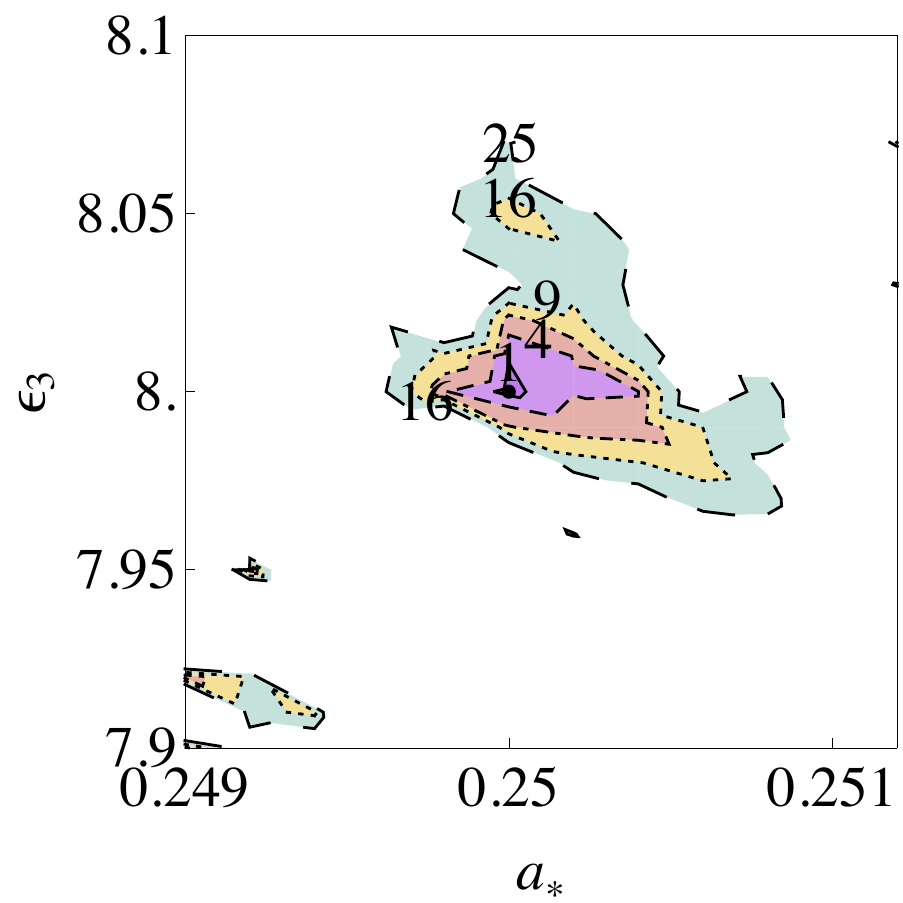}
\end{center}
\vspace{-0.5cm}
\caption{Top panels: $\Delta \chi_{\rm r}^2$ contours with 
$N=10^3$ from the comparison of the 2D transfer 
function of a Johannsen-Psaltis BH with a spin parameter $a_*' = 0.25$, deformation 
parameter $\epsilon_3' = 8$, and an inclination of $i'=45^\circ$ versus a set of 
Johannsen-Psaltis BHs with spin parameters $a_*$ ($x$ axis) and deformation 
parameters $\epsilon_3$ ($y$ axis). In the left panel, all models use an inclination 
angle $i'= i = 45^\circ$. In the right panel, $i$ is free in the fit. For the sake of simplicity, the emissivity index is 
fixed to $q' = q = 3$ and the source height is $h' = h = 10 \, M$. The time range 
of the 2D transfer function is $[0, 300 \, M]$. Bottom panels: as in the top panels 
with $i = 45^\circ$ (left panel) and $i$ free (right panel) for $N=10^5$. See the 
text for more details.}
\label{fig17}
\end{figure}

\subsection{A non-Kerr black hole example \label{ss-ex}}

Thus far, we have broadly explored a select few correlations within a large 
multi-dimensional parameter space.  We now more carefully consider one
particular non-Kerr template model to assess whether we can constrain the 
deformation parameter 
and whether we can distinguish our object from a Kerr BH. 
We assume a Johannsen-Psaltis BH
with spin parameter $a_*' = 0.25$ and deformation parameter $\epsilon_3' = 8$.
We similarly adopt an inclination angle $i' = 45^\circ$, an emissivity index $q'
= 3$, and a height for the X-ray source $h' = 10 \, M$, and use this
as our reference model.

First, we consider the time-integrated line profile.  
Fig.~\ref{fig16} shows contours of $\Delta\chi^2_{\rm p}$ with 
$N=10^3$ (top panels) and $10^5$ (bottom panels). 
In the left panels, we make the simplifying assumption that the inclination angle 
and the emissivity index are known, i.e., we set $i = i'$ and $q = q'$. In the right
panels, $i$ is instead treated as a fit parameter.  To isolate the importance of inclination, we again adopt $q = q'$.
However, such an approach is motivated given the weak correlation between 
$q$ and the geometry parameters, as discussed above. 
We note a few important points. First, for $N=10^3$
if the inclination angle were known 
then the iron line spectrum could on its own establish the non-vanishing deformation 
parameter of our template robustly (top left panel in Fig.~\ref{fig16}). However,
the power of this constraint is substantially reduced  when allowing inclination to vary.  Second, for $N=10^5$ we have a substantial 
improvement in the measurement quality allowing for strong testing of the Kerr metric (bottom panels in Fig.~\ref{fig16}). Third,
in the case of $N=10^5$ we break the degeneracy arising from the
inclination angle.

A parallel analysis has been performed but considering the full reverberation
mapping signal, as opposed to the integrated line profile.  
The results are shown in Fig.~\ref{fig17}, for $N=10^3$ (top panels) 
and $N=10^5$ (bottom panels). As in Fig.~\ref{fig16}, for the left panels, $i = i'$, 
while in the right panels $i$ is treated as a fit parameter and we have minimized 
$\chi^2_{\rm r}$ across all inclination values. As discussed above, we
have adopted $q = q' = 3$, $h = h' = 10 \, M$, and the time range is $[0,300\,M]$. 
Reverberation mapping is more powerful than the time-averaged iron line profile, and the difference is most pronounced for the constraint on spin.   If we compare the top right panels in Figs.~\ref{fig16} and 
\ref{fig17}, the allowed regions are narrower in spin for the case of reverberation mapping, 
but it is difficult to constrain a Kerr deviation with $N = 10^3$ photons. But we find the situation is much improved with sufficiently high signal.  
An observation
with $N = 10^5$ photons can readily distinguish a Kerr BH from one with 
$\epsilon_3 = 8$ from the integrated spectrum alone.  However, the time information improves the quality of constraint by roughly an order of magnitude, from $\Delta \epsilon_3 \sim 1$ (profile), to $\Delta \epsilon_3 \sim 0.1$ (reverberation).  This enormous gain means a much more potent constraint on deviations from Kerr is enabled by additional time information, but clearly also requires appreciable signal to make the difference.

While here we have provided just a single example with a large value of the deformation parameter, the behavior of the metric with respect to $\epsilon_3$ is relatively smooth. As a crude estimate, we can imagine the constraint for a different value of the spin and the deformation parameter of the reference model simply adopting the uncertainty on $a_*$ and $\epsilon_3$ shown in Figs.~\ref{fig16} and \ref{fig17}.

\section{Discussion \label{s-d}}

In the previous section, we compared a number of Kerr and 
non-Kerr models to determine if and how reverberation measurements  
provide extra information about the nature of BH candidates compared
to the readily available measurements of time-integrated iron line spectra. Within our 
simplified analysis, we have
used $\chi^2_{\rm p} \approx N \mathcal{L}_{\rm p}$ and $\chi^2_{\rm r} \approx 
N \mathcal{L}_{\rm r}$ to quantify the level of similarity between iron line profiles 
and between 2D reverberation transfer functions for different models.
Here $N$ is the total photon number and $\mathcal{L}_{\rm p,r}$ is the
log-likelihood defined in Eqs.~(\ref{log-l1}) and (\ref{log-l2}). In our figures,
we have assumed either $N = 10^3$ or $N = 10^5$. At first approximation, our 
results can be rescaled for a different photon number, but the fact we have 
included the photon noise and required a minimum of 10~photons per bin
introduces some deviations from an exact scaling.

Although we have adopted several simplifications, our results indicate
that an accurate measurement of the 2D transfer function has advantages
for testing the Kerr nature of astrophysical BH candidates compared to 
the analysis of the iron line profile alone. 
However, the difference is only substantial at high photon count; 
$N = 10^5$ photons can strongly constrain the Kerr metric even
in the case of a time-integrated measurement. It should be clear that actually
the constraints shown in the bottom panels in Fig.~\ref{fig16} and \ref{fig17}
are not realistic as at a certain point systematics effects due to a simplified
model become important. Despite that, from Fig.~\ref{fig16} and \ref{fig17}
it is clear that current iron line data cannot rule out even large deviations from 
the Kerr solution, even in the best cases. The possibility 
of future detections with a large number of photons in the iron line can provide 
interesting constraints and test the Kerr paradigm. Much of the information
on the spacetime geometry is already encoded in the spectrum alone (e.g., \cite{cackett}). However, reverberation 
mapping is potentially a more reliable technique to understand the systematics. 
With added time information, we can distinguish photons emitted at different radii 
and therefore affected in a different way by relativistic effects. In the case of the 
time-integrated iron line, photons emitted from different radii can sum up in the 
same energy bin and so added the information is lost. 
For example, a background independent estimate of the inclination angle from 
the late time 2D transfer function is a robustly determined quality.

Lastly, we note that our theoretical model, and this investigation, employ numerous 
simplifications. While these simplifications are reasonable for an exploration,
followup study using a more thorough treatment of the
various model degeneracies is necessary before a serious
observational campaign for testing the Kerr BH hypothesis would be
viable.   There are many assumptions made here, and even elsewhere in
iron-line spectral modeling that should be improved upon (and indeed,
are the subject of ongoing work).  
For example, here the flaring region has been
approximated by an on-axis point-like source. A more realistic model should be able to
take into account multiple locations, extended geometry, or motion of
the illuminating  X-ray source~\cite{wil12}. The emissivity function has been modeled as
a power law $\sim 1/r^q$ with constant emissivity index $q$, but even in the simple
lamppost geometry there are significant departures at small radii, where the information
about the spacetime geometry come from~\cite{dauser}.  Meanwhile, at
large scales there can be a variety of
more subtle astrophysical complications, including the possibility of
disk warps or hotspots.  Multiple ionization states, and a more
detailed spectral modeling going beyond the iron line to consider the
full reflection spectrum would also be important to consider (e.g., \cite{gar1}). 
The impact of all these effects should be assessed. Here, our
goal is to provide a first exploratory step along this path, in order
to illuminate the most prominent effects and pave the way for
followup studies.

\section{Summary and conclusions \label{s-c}}

Astrophysical BH candidates are supposed to be the Kerr BHs predicted in general 
relativity, but the actual nature of these objects has still to be confirmed. The study 
of the properties of the electromagnetic radiation emitted by the gas in the accretion 
disk can be used to probe the spacetime geometry around these compact objects
and hopefully test the Kerr BH hypothesis. Today, the continuum-fitting method and 
the analysis of the iron K$\alpha$ line are the primary techniques capable of providing 
information about the metric around BH candidates. However, it is usually difficult to
simultaneously constrain possible deviations from the Kerr solution, because there 
is a strong correlation between the measurement of the spin and the deformation 
parameters.  In the end, non-Kerr BHs may look like Kerr BHs with a different spin 
parameter and, without an independent estimate of its spin, it is impossible to test their Kerr nature.

The iron K$\alpha$ line is produced from the illumination of a cold disk by a hot corona 
above the accretion disk.  Flaring regions in the corona produce fast time variability 
in the fluorescent line emission in the disk, manifesting in line reverberation. Because 
of insufficient count rates in present X-ray facilities in the iron-line band, to achieve 
sufficient signal one usually integrates for thousands of seconds.  This causes a 
loss of information. Future X-ray facilities with larger effective areas may be capable of 
measuring the 2D transfer function; that is, the iron line signal as a function of time in 
response to an instantaneous coronal flare. The determination of the 2D transfer function 
can provide information about the spacetime geometry around the BH candidate, the 
geometry of the primary X-ray source, and the geometry of the accretion disk.

In this Paper, we report an exploratory and preliminary study on the
impact of deviations from Kerr geometry 
on iron-line reverberation.  We accordingly make simplistic and
approximate predictions at the prospect of
using iron-line reverberation to test the Kerr nature of BH candidates.  
In Section~\ref{s-sim}, we have considered a few specific examples for
which we compare iron  line profiles and 2D transfer functions in Kerr
versus non-Kerr backgrounds. 
With the caution that we have adopted a simplified theoretical framework and not 
attempted to fully explore the correlations between all model parameters, our 
results indicate that 
reverberation measurements can provide stronger
constraints on the spacetime geometry around astrophysical 
BH candidates with respect to time-integrated data. However, the
difference is only substantial for high-signal data. 
With available X-ray facilities, even a good observation with $\sim 10^3$
photons in the iron line cannot distinguish a Kerr BH from a very deformed
object. 
As an optimistic future observation, we have considered the case $N \sim10^5$ and found that a similar observation can test the Kerr metric. From a simple interpolation between the case $N \sim10^3$ and $N \sim10^5$, we may expect that $\sim10^4$ photons could already provide interesting constraints. Such results will hopefully
be achieved in the near future with a generational improvement in X-ray instruments.

Although reverberation studies are already being used to glean information about 
relativistic effects in the inner reaches of AGN systems (see e.g 
Refs.~\cite{zog2010,zog2012,kara2013,2-emman}), for the precision measurements 
we consider, the present fleet of X-ray detectors are unable to adequately resolve 
the transfer function at the required  time plus energy resolution (they fall short of the 
mark for both stellar-mass and supermassive BH candidates). 
While a more detailed analysis beyond the scope of this work is necessary to properly 
assess the observability of non-Kerr deviations in  reverberation data,  
our exploratory results indicate that a next-generation successor to
{\it RXTE}~\cite{swank99} (such as 
{\it LOFT}~\cite{fero12} and possibly Athena~\cite{v4-athena}) might be capable of producing such measurements for a 
select few of the brightest and nearest AGN systems.

For instance, for {\it LOFT} observations of a 
nearby AGN, 
$z<0.1$ with mass $\sim10^9 M_\odot$,
one may achieve {\em thousands} of photon counts in its 
reflection features, per unit $M$ in time (i.e., bins of $\sim 5000$s). The most 
prominent of these reflection features is that which we have emphasized throughout, 
namely the iron K$\alpha$ line. 
Of course, in practice one may take advantage of signal from the {\em full} reflection spectrum, beyond just 
the iron line.  
Even while this domain becomes accessible to us for supermassive BH candidate 
systems, the timescales are sufficiently short for stellar-mass systems (i.e, tens of 
$\mu$s), that no detector now or on the immediate horizon, is likely to allow 
measurement of a transfer function in the relativistic regime for stellar-mass BH 
candidates.


\begin{acknowledgments}
We would like to thank Gary Hinshaw for useful comments
and suggestions.
JJ and CB were supported by the NSFC grant No.~11305038,
the Shanghai Municipal Education Commission grant for Innovative
Programs No.~14ZZ001, the Thousand Young Talents Program,
and Fudan University.
JFS was supported by the NASA Hubble Fellowship grant
HST-HF-51315.01.
\end{acknowledgments}



\begin{thebibliography}{99}

\bibitem{h1}
  B.~Carter,
  Phys.\ Rev.\ Lett.\  {\bf 26}, 331 (1971).

\bibitem{h2}
  D.~C.~Robinson,
  Phys.\ Rev.\ Lett.\  {\bf 34}, 905 (1975).

\bibitem{h3}
  P.~T.~Chrusciel, J.~L.~Costa and M.~Heusler,
  Living Rev.\ Rel.\  {\bf 15}, 7 (2012)
  [arXiv:1205.6112 [gr-qc]].

\bibitem{nara}
  R.~Narayan,
  New J.\ Phys.\  {\bf 7}, 199 (2005)
  [gr-qc/0506078].

\bibitem{p1}
  R.~H.~Price,
  Phys.\ Rev.\ D {\bf 5}, 2419 (1972).

\bibitem{p2}
  R.~H.~Price,
  Phys.\ Rev.\ D {\bf 5}, 2439 (1972).

\bibitem{bdp}
  C.~Bambi, A.~D.~Dolgov and A.~A.~Petrov,
  JCAP {\bf 0909}, 013 (2009)
  [arXiv:0806.3440 [astro-ph]].
  
\bibitem{massivedisk}
  C.~Bambi, D.~Malafarina and N.~Tsukamoto,
  Phys.\ Rev.\ D {\bf 89}, 127302 (2014)
  [arXiv:1406.2181 [gr-qc]].

\bibitem{ozel1}
F.~{\"O}zel, D.~Psaltis, R.~Narayan and J.~E.~McClintock,
  Astrophys.\ J.\  {\bf 725}, 1918 (2010)
  [arXiv:1006.2834 [astro-ph.GA]].

\bibitem{ozel2}
  F.~{\"O}zel, G.~Baym and T.~Guver,
  Phys.\ Rev.\ D {\bf 82}, 101301 (2010)
  [arXiv:1002.3153 [astro-ph.HE]].

\bibitem{bh1}
  C.~E.~Rhoades and R.~Ruffini,
  Phys.\ Rev.\ Lett.\  {\bf 32}, 324 (1974).

\bibitem{bh2}
  V.~Kalogera and G.~Baym,
  Astrophys.\ J.\  {\bf 470}, L61 (1996).

\bibitem{bh3}
  E.~Maoz,
  Astrophys.\ J.\  {\bf 494}, L181 (1998)
  [astro-ph/9710309].

\bibitem{bh4}
  R.~Narayan and J.~E.~McClintock,
  New Astron.\ Rev.\  {\bf 51}, 733 (2008)
  [arXiv:0803.0322 [astro-ph]].

\bibitem{bh5}
  A.~E.~Broderick, A.~Loeb and R.~Narayan,
  Astrophys.\ J.\  {\bf 701}, 1357 (2009)
  [arXiv:0903.1105 [astro-ph.HE]].

\bibitem{will}
  C.~M.~Will,
  Living Rev.\ Rel.\  {\bf 9}, 3 (2006)
  [gr-qc/0510072].

\bibitem{gia1}
  G.~Dvali and C.~Gomez,
  Fortsch.\ Phys.\  {\bf 61}, 742 (2013)
  [arXiv:1112.3359 [hep-th]].  
  
\bibitem{gia2}
  G.~Dvali and C.~Gomez,
  Phys.\ Lett.\ B {\bf 719}, 419 (2013)
  [arXiv:1203.6575 [hep-th]].  
  
\bibitem{giddings}
  S.~B.~Giddings,
  Phys.\ Rev.\ D {\bf 90}, no. 12, 124033 (2014)
  [arXiv:1406.7001 [hep-th]].  

\bibitem{gw}
  F.~D.~Ryan,
  Phys.\ Rev.\  D {\bf 52}, 5707 (1995).

\bibitem{gw1}
  K.~Glampedakis and S.~Babak,
  Class.\ Quant.\ Grav.\  {\bf 23}, 4167 (2006)
  [arXiv:gr-qc/0510057].

\bibitem{gw2}
  L.~Barack and C.~Cutler,
  Phys.\ Rev.\  D {\bf 75}, 042003 (2007)
  [arXiv:gr-qc/0612029].

\bibitem{gw3}
  T.~A.~Apostolatos, G.~Lukes-Gerakopoulos and G.~Contopoulos,
  Phys.\ Rev.\ Lett.\  {\bf 103}, 111101 (2009)
  [arXiv:0906.0093 [gr-qc]].

\bibitem{cfm1}
  D.~F.~Torres,
ÊÊNucl.\ Phys.\ B {\bf 626}, 377 (2002)
ÊÊ[hep-ph/0201154].

\bibitem{cfm2}
  C.~Bambi and E.~Barausse,
  Astrophys.\ J.\  {\bf 731}, 121 (2011)
  [arXiv:1012.2007 [gr-qc]].

\bibitem{cfm3}
  C.~Bambi,
  Astrophys.\ J.\  {\bf 761}, 174 (2012)
  [arXiv:1210.5679 [gr-qc]].

\bibitem{cfm4}
  L.~Kong, Z.~Li and C.~Bambi,
  Astrophys.\ J.\  {\bf 797}, 78 (2014)
  [arXiv:1405.1508 [gr-qc]].  

\bibitem{iron1}
  Y.~Lu and D.~F.~Torres,
ÊÊInt.\ J.\ Mod.\ Phys.\ D {\bf 12}, 63 (2003)
ÊÊ[astro-ph/0205418].

\bibitem{iron2}
  T.~Johannsen and D.~Psaltis,
  Astrophys.\ J.\  {\bf 773}, 57 (2013)
  [arXiv:1202.6069 [astro-ph.HE]].

\bibitem{iron3}
  C.~Bambi,
  Phys.\ Rev.\ D {\bf 87}, 023007 (2013)
  [arXiv:1211.2513 [gr-qc]].

\bibitem{qpo1}
  T.~Johannsen and D.~Psaltis,
  Astrophys.\ J.\  {\bf 726}, 11 (2011)
  [arXiv:1010.1000 [astro-ph.HE]].

\bibitem{qpo2}
  C.~Bambi,
  JCAP {\bf 1209}, 014 (2012)
  [arXiv:1205.6348 [gr-qc]].

\bibitem{qpo3}
  A.~N.~Aliev, G.~D.~Esmer and P.~Talazan,
  Class.\ Quant.\ Grav.\  {\bf 30}, 045010 (2013)
  [arXiv:1205.2838 [gr-qc]].

\bibitem{cfm-iron}
  C.~Bambi,
  JCAP {\bf 1308}, 055 (2013)
  [arXiv:1305.5409 [gr-qc]].

\bibitem{sh1}
  C.~Bambi and K.~Freese,
  Phys.\ Rev.\ D {\bf 79}, 043002 (2009)
  [arXiv:0812.1328 [astro-ph]].

\bibitem{sh2}
  C.~Bambi and N.~Yoshida,
  Class.\ Quant.\ Grav.\  {\bf 27}, 205006 (2010)
  [arXiv:1004.3149 [gr-qc]].

\bibitem{sh3}
  C.~Bambi,
  Phys.\ Rev.\ D {\bf 87}, 107501 (2013)
  [arXiv:1304.5691 [gr-qc]].

\bibitem{sh4}
  Z.~Li and C.~Bambi,
  JCAP {\bf 1401}, 041 (2014)
  [arXiv:1309.1606 [gr-qc]].

\bibitem{pulsar}
  K.~Liu, N.~Wex, M.~Kramer, J.~M.~Cordes and T.~J.~W.~Lazio,
  Astrophys.\ J.\  {\bf 747}, 1 (2012)
  [arXiv:1112.2151 [astro-ph.HE]].

\bibitem{hot2}
  Z.~Li and C.~Bambi,
  Phys.\ Rev.\ D {\bf 90}, 024071 (2014)
  [arXiv:1405.1883 [gr-qc]].
  
\bibitem{2-aaa} 
  F.~Atamurotov, A.~Abdujabbarov and B.~Ahmedov,
  Phys.\ Rev.\ D {\bf 88}, 064004 (2013).

\bibitem{rev1}
  C.~Bambi,
  Mod.\ Phys.\ Lett.\ A {\bf 26}, 2453 (2011)
  [arXiv:1109.4256 [gr-qc]].

\bibitem{rev2}
  C.~Bambi,
  Astron.\ Rev.\  {\bf 8}, 4 (2013)
  [arXiv:1301.0361 [gr-qc]].

\bibitem{k1}
  S.~N.~Zhang, W.~Cui and W.~Chen,
  Astrophys.\ J.\  {\bf 482}, L155 (1997)
  [astro-ph/9704072].

\bibitem{k2}
  L.~-X.~Li, E.~R.~Zimmerman, R.~Narayan and J.~E.~McClintock,
  Astrophys.\ J.\ Suppl.\  {\bf 157}, 335 (2005)
  [astro-ph/0411583].

\bibitem{k3}
  J.~E.~McClintock, R.~Narayan, S.~W.~Davis, L.~Gou, A.~Kulkarni, J.~A.~Orosz, R.~F.~Penna and R.~A.~Remillard {\it et al.},
  Class.\ Quant.\ Grav.\  {\bf 28}, 114009 (2011)
  [arXiv:1101.0811 [astro-ph.HE]].

\bibitem{k3b}
  J.~E.~McClintock, R.~Narayan and J.~F.~Steiner,
  Space~Science~Reviews 73 (2014)
  [arXiv:1303.1583 [astro-ph.HE]]. 

\bibitem{k4}
  A.~C.~Fabian, M.~J.~Rees, L.~Stella and N.~E.~White,
  Mon.\ Not.\ Roy.\ Astron.\ Soc.\  {\bf 238}, 729 (1989).

\bibitem{k5}
  A.~C.~Fabian, K.~Iwasawa, C.~S.~Reynolds and A.~J.~Young,
  Publ.\ Astron.\ Soc.\ Pac.\  {\bf 112}, 1145 (2000)
  [astro-ph/0004366].

\bibitem{k6}
  L.~Brenneman,
  {\it Measuring the Angular Momentum of Supermassive Black Holes},
  (SpringerBriefs in Astronomy, ISBN 978-1-4614-7770-9, 2013).

\bibitem{k6b}
  C.~S.~Reynolds,
  Space~Science~Reviews 81
  [arXiv:1302.3260 [astro-ph.HE]].

\bibitem{ste1}
  J.~F.~Steiner, J.~E.~McClintock, R.~A.~Remillard, L.~Gou, S.~'y.~Yamada and R.~Narayan,
  Astrophys.\ J.\  {\bf 718}, L117 (2010)
  [arXiv:1006.5729 [astro-ph.HE]].

\bibitem{vnnn-1} 
  J.~E.~McClintock, R.~Shafee, R.~Narayan, R.~A.~Remillard, S.~W.~Davis and L.~X.~Li,
  Astrophys.\ J.\  {\bf 652}, 518 (2006)
  [astro-ph/0606076].

\bibitem{vnnn-2} 
  A.~Sadowski, M.~Abramowicz, M.~Bursa, W.~Kluzniak, J.~P.~Lasota and A.~Rozanska,
  Astron.\ Astrophys.\  {\bf 527}, A17 (2011)
  [arXiv:1006.4309 [astro-ph.HE]].

\bibitem{v4-m1} 
  S.~E.~Motta, T.~M.~Belloni, L.~Stella, T.~Mu–oz-Darias and R.~Fender,
  Mon.\ Not.\ Roy.\ Astron.\ Soc.\  {\bf 437}, 2554 (2014)
  [arXiv:1309.3652 [astro-ph.HE]].

\bibitem{v4-m2} 
  S.~E.~Motta, T.~Mu–oz-Darias, A.~Sanna, R.~Fender, T.~Belloni and L.~Stella,
  Mon.\ Not.\ Roy.\ Astron.\ Soc.\  {\bf 439}, 65 (2014)
  [arXiv:1312.3114 [astro-ph.HE]].

\bibitem{v4-r} 
  G.~Risaliti, E.~Nardini, M.~Elvis, L.~Brenneman and M.~Salvati,
  Mon.\ Not.\ Roy.\ Astron.\ Soc.\  {\bf 417}, 178 (2011)
  [arXiv:1105.2318 [astro-ph.CO]].

\bibitem{v4-matt} 
  M.~Middleton and A.~Ingram,
  Mon.\ Not.\ Roy.\ Astron.\ Soc.\  {\bf 446}, 1312 (2015)
  [arXiv:1410.5992 [astro-ph.HE]].
  
\bibitem{v4-cb}
  C.~Bambi,
  arXiv:1312.2228 [gr-qc].

\bibitem{j1}
  C.~Bambi,
  Phys.\ Rev.\ D {\bf 85}, 043002 (2012)
  [arXiv:1201.1638 [gr-qc]].

\bibitem{j2}
  C.~Bambi,
  Phys.\ Rev.\ D {\bf 86}, 123013 (2012)
  [arXiv:1204.6395 [gr-qc]].

\bibitem{hot1}
  Z.~Li, L.~Kong and C.~Bambi,
  Astrophys.\ J.\  {\bf 787}, 152 (2014)
  [arXiv:1401.1282 [gr-qc]].

\bibitem{naoki}
  N.~Tsukamoto, Z.~Li and C.~Bambi,
  JCAP {\bf 1406}, 043 (2014)
  [arXiv:1403.0371 [gr-qc]].

\bibitem{cc1}
C.~Bambi and D.~Malafarina,
  Phys.\ Rev.\ D {\bf 88}, 064022 (2013)
  [arXiv:1307.2106 [gr-qc]].

\bibitem{cc2}
  C.~Bambi,
  Phys.\ Rev.\ D {\bf 87}, 084039 (2013)
  [arXiv:1303.0624 [gr-qc]].

\bibitem{cc3}
  C.~Bambi,
  Phys.\ Lett.\ B {\bf 730}, 59 (2014)
  [arXiv:1401.4640 [gr-qc]].
  
\bibitem{cpr-metric}
  C.~Bambi,
  Phys.\ Rev.\ D {\bf 90}, 047503 (2014)
  [arXiv:1408.0690 [gr-qc]].

\bibitem{v4-wg} 
  D.~R.~Wilkins and L.~C.~Gallo,
  Mon.\ Not.\ Roy.\ Astron.\ Soc.\  {\bf 448}, 703 (2015)
  [arXiv:1412.0015 [astro-ph.HE]].

\bibitem{v4-uttley} 
  P.~Uttley, I.~M.~McHardy and S.~Vaughan,
  Mon.\ Not.\ Roy.\ Astron.\ Soc.\  {\bf 359}, 345 (2005)
  [astro-ph/0502112].

\bibitem{r1}
  L.~Stella,
  Nature {\bf 344}, 747 (1990).

\bibitem{r2}
  G.~Matt and C.~Perola,
  Mon.\ Not.\ Roy.\ Astron.\ Soc.\  {\bf 259}, 433 (1992).

\bibitem{r3}
  C.~S.~Reynolds, A.~J.~Young, M.~C.~Begelman and A.~C.~Fabian,
  Astrophys.\ J.\  {\bf 514}, 164 (1999)
  [astro-ph/9806327].

\bibitem{l1}
  G.~Matt, G.~C.~Perola and L.~Piro,
  Astron.\ Astrophys.\  {\bf 247}, 25 (1991).

\bibitem{l2}
  A.~Martocchia and G.~Matt,
  Mon.\ Not.\ Roy.\ Astron.\ Soc.\  {\bf 282}, L53 (1996).

\bibitem{v4-sw1} 
  B.~Czerny and A.~Janiuk,
ÊÊAstron.\ Astrophys.\  {\bf 464}, 167 (2007)
ÊÊ[astro-ph/0612262].

\bibitem{v4-sw2} 
  J.~D.~Schnittman and J.~H.~Krolik,
ÊÊAstrophys.\ J.\  {\bf 712}, 908 (2010)
ÊÊ[arXiv:0912.0907 [astro-ph.HE]].

\bibitem{v4-sw3} 
  C.~Done, S.~Davis, C.~Jin, O.~Blaes and M.~Ward,
  Mon.\ Not.\ Roy.\ Astron.\ Soc.\  {\bf 420}, 1848 (2012)
  [arXiv:1107.5429 [astro-ph.HE]].
  
\bibitem{nt-model}
  I.~D.~Novikov, K.~S.~Thorne,
  ``Astrophysics of Black Holes'' in {\it Black Holes}, 
  edited by C.~De~Witt and B.~De~Witt
  (Gordon and Breach, New York, New York, 1973), pp. 343-450.
  
\bibitem{nt-model2}
  D.~N.~Page and K.~S.~Thorne,
  Astrophys.\ J.\  {\bf 191}, 499 (1974). 

\bibitem{gfab}
    I.~M.~George, A.~C.~Fabian,
   Mon.\ Not.\ Roy.\ Astron.\ Soc.\  {\bf 249}, 352 (1991).


\bibitem{rofab}
    R.~R.~Ross, A.~C.~Fabian,
   Mon.\ Not.\ Roy.\ Astron.\ Soc.\  {\bf 358}, 211 (2005)
  [arXiv: 0501116 [astro-ph]]. 
   


\bibitem{gar1}
  J.~Garcia, T.~Dauser, A.~Lohfink, T.~R.~Kallman, J.~F.~Steiner, J.~E.~McClintock, L.~Brenneman and J.~Wilms {\it et al.},
  Astrophys.\ J.\  {\bf 782}, 76 (2014)
  [arXiv:1312.3231 [astro-ph.HE]].

\bibitem{dauser}
  T.~Dauser, J.~Garcia, J.~Wilms, M.~Bock, L.~W.~Brenneman, M.~Falanga, K.~Fukumura and C.~S.~Reynolds,
  Mon.\ Not.\ Roy.\ Astron.\ Soc.\  {\bf 430}, 1694 (2013)
  [arXiv:1301.4922 [astro-ph.HE]].


\bibitem{king05}
   A.~R.~King, S.~H.~Lubow, G.~I.~Ogilvie, and J.~E.~Pringle,
  Mon.\ Not.\ Roy.\ Astron.\ Soc.\  {\bf 363}, 49 (2005)
  [arXiv: 0507098 [astro-ph]]. 


\bibitem{zhu1}
  Y.~Zhu, S.~W.~Davis, R.~Narayan, A.~K.~Kulkarni, R.~F.~Penna and J.~E.~McClintock,
  Mon.\ Not.\ Roy.\ Astron.\ Soc.\  {\bf 424}, 2504 (2012)
  [arXiv:1202.1530 [astro-ph.HE]]. 

\bibitem{schnit1} 
  J.~D.~Schnittman, J.~H.~Krolik and S.~C.~Noble,
  Astrophys.\ J.\  {\bf 769}, 156 (2013)
  [arXiv:1207.2693 [astro-ph.HE]].

\bibitem{jpm}
  T.~Johannsen and D.~Psaltis,
  Phys.\ Rev.\ D {\bf 83}, 124015 (2011)
  [arXiv:1105.3191 [gr-qc]].
  
\bibitem{petebook}
  B.~M.~Peterson,
  {\it An introduction to active galactic nuclei} 
  (Cambridge University Press, Cambridge, UK, 1997).
 
\bibitem{cackett}
  E.~M.~Cackett, A.~Zoghbi, C.~Reynolds, A.~C.~Fabian, E.~Kara, P.~Uttley and D.~R.~Wilkins,   
  Mon.\ Not.\ Roy.\ Astron.\ Soc.\  {\bf 438}, 2980 (2014)
  [arXiv:1311.2997 [astro-ph.HE]].  

\bibitem{wil12} 
  D.~R.~Wilkins and A.~C.~Fabian,
  Mon.\ Not.\ Roy.\ Astron.\ Soc.\  {\bf 424}, 1284 (2012)
  [arXiv:1205.3179 [astro-ph.HE]].
  
\bibitem{zog2010} 
  A.~Zoghbi, A.~Fabian, P.~Uttley, G.~Miniutti, L.~Gallo, C.~Reynolds, J.~Miller and G.~Ponti,
  Mon.\ Not.\ Roy.\ Astron.\ Soc.\  {\bf 401}, 2419 (2010)
  [arXiv:0910.0367 [astro-ph.HE]].

\bibitem{zog2012} 
  A.~Zoghbi, A.~C.~Fabian, C.~S.~Reynolds and E.~M.~Cackett,
  Mon.\ Not.\ Roy.\ Astron.\ Soc.\  {\bf 422}, 129 (2012)
  [arXiv:1112.1717 [astro-ph.HE]].

\bibitem{kara2013} 
  E.~Kara, A.~C.~Fabian, E.~M.~Cackett, J.~F.~Steiner, P.~Uttley, D.~R.~Wilkins and A.~Zoghbi,
  Mon.\ Not.\ Roy.\ Astron.\ Soc.\  {\bf 428}, 2795 (2013)
  [arXiv:1210.1465 [astro-ph.HE]].
  
\bibitem{2-emman} 
  D.~Emmanoulopoulos, I.~E.~Papadakis, M.~Dovciak and I.~M.~McHardy,
  Mon.\ Not.\ Roy.\ Astron.\ Soc.\  {\bf 439}, 3931 (2014)
  [arXiv:1402.0899 [astro-ph.HE]].

\bibitem{swank99}
  J.~H.~Swank,
  Nucl.\ Phys.\ B Proc. Suppl. {\bf 69}, 12 (1999)
  [astro-ph/9802188].

\bibitem{fero12}
  M.~Feroci, L.~Stella and M.~van der Klis {\it et al.},
  Experimental Astronomy {\bf 34}, 415 (2012).
  
\bibitem{v4-athena} 
  K.~Nandra, D.~Barret, X.~Barcons, A.~Fabian, J.~W.~d.~Herder, L.~Piro, M.~Watson and C.~Adami {\it et al.},
  arXiv:1306.2307 [astro-ph.HE].  

\end{thebibliography}
\end{document}